\documentclass[aps,prb,twocolumn,floatfix,superscriptaddress,showpacs,footinbib]{revtex4}
\usepackage{amsmath,amssymb,natbib,bm,graphicx,url,epsfig}
\usepackage{ulem}
\usepackage[ansinew]{inputenc}

\usepackage{color}

\begin{document}

\vspace*{0.8ex}

\title{Resonant and Kondo tunneling through molecular magnets}

\author{Florian Elste}

\affiliation{Department of Physics, Columbia University, 538 West
120th Street, New York, New York 10027, USA}

\author{Carsten Timm}

\affiliation{Institut f\"ur Theoretische Physik, Technische
Universit\"at Dresden, 01062 Dresden, Germany}

\date{\today}

\begin{abstract}
Transport through molecular magnets is studied in the regime of strong
coupling to the leads. We consider a resonant-tunneling model where the
electron spin in a quantum dot or molecule is coupled to an additional local,
anisotropic spin via exchange interaction. The two opposite regimes dominated by
resonant tunneling and by Kondo transport, respectively, are considered. In
the resonant-tunneling regime, the stationary state of the impurity spin is
calculated for arbitrarily strong molecule-lead coupling using a master-equation
approach, which treats the exchange interaction perturbatively. We find that
the characteristic fine structure in the differential conductance persists even
if the hybridization energy exceeds thermal energies. Transport in the Kondo
regime is studied within a diagrammatic approach. 
We show that magnetic anisotropy gives rise to the appearance of two Kondo
peaks at nonzero bias voltages. 
\end{abstract}


\maketitle

\section{Introduction}

Over the past few years the idea of integrating the concepts of spintronics and
molecular electronics has developed into a new research field dubbed
\textit{molecular spintronics}.\cite{Sanvito,Bogani} Progress has not only
been stimulated by technological interests but has also been accompanied by
the realization that magnetic single-molecule transistors exhibit various
fundamental quantum phenomena.\cite{Park,Jo,Heersche,Grose,Tejada,Durkan,Rugar}
Among many promising ideas discussed in the literature, particular attention has
been paid to current-induced spin reading and writing, spin relaxation,
entanglement, quantum computation, and Kondo
correlations.\cite{Romeike,Elste,Misiorny1,Misiorny2,Braun,%
Barraza-Lopez,Souza,Jonckheere,Lu,Lindebaum}

An experimental realization of spintronics devices may be achieved by using
single-molecule magnets in combination with metallic (nonmagnetic or
ferromagnetic) leads. For molecular-memory applications, long spin-relaxation
times are advantageous, which may be realized in molecules with large magnetic
anisotropy, such as molecules based on $\mathrm{Mn}_{12}$, $\mathrm{Fe}_4$,
and $\mathrm{Ni}_4$.\cite{Blundell,Sangregorio,Mannini}

Controlling and detecting the molecular spin by means of electronic tunneling
into source and drain electrodes poses a major challenge. While some approaches
rely on break junctions, others are based on a scanning tunneling microscope.
In both cases, the coupling between the molecule and the leads can vary by
several orders of magnitude, thus giving rise to strikingly different transport
regimes.\cite{Park,Jo,Heersche,Grose,Nitzan,Xue,Joachim,Donarini,%
Zhou,Yu,Emberly2,Koch,Fang,Koerting}

In the regime of weak molecule-lead coupling, many experimental features such as
Coulomb blockade, spin blockade, sequential tunneling, and cotunneling can
be described within a master-equation or rate-equation approach treating the
electronic tunneling perturbatively.\cite{Timm} However, for strong
coupling, low-order perturbation theory breaks down. It is then advantageous to
treat the electronic tunneling exactly, at the price of introducing
approximations elsewhere.

Recently, the Kondo effect in single-molecule magnets with easy-axis anisotropy
has been studied by Romeike \textit{et al.}\cite{Romeike} Their model
describes an anisotropic spin coupled to metallic electrodes by an
exchange interaction, in the absence of a bias voltage. This differs from the
model studied here, in which the
electronic spin in the relevant molecular orbital is coupled to an additional
local, anisotropic spin via an exchange interaction $J$, i.e., charge
fluctuations are explicitly taken into account. In addition, we include a
nonzero bias voltage. The presence of a
Kondo effect for an anisotropic spin is at first glance surprising since
it requires two approximately degenerate low-energy spin states connected by a
term in the Hamiltonian. The simplest Hamiltonian for an anisotropic spin
$\mathbf{S}$ with $S\ge 1$ exchange-coupled to an electronic spin $\mathbf{s}$
\begin{equation}
H = -K_2 (S^z)^2 + J\, \mathbf{s}\cdot\mathbf{S}
\label{basicH}
\end{equation}
does not provide such a term. Using a renormalization-group approach, Romeike
\textit{et al.}\cite{Romeike} could show that quantum tunneling of the magnetic
moment, which is described by higher-order anisotropy terms not included in
Eq.\ (\ref{basicH}), may give rise to a Kondo peak in the linear
conductance, centered at zero bias voltage. The Kondo temperature is found
to depend strongly on the ratio of the applied magnetic field and the anisotropy
barrier. Further, Gonz\'alez~\textit{et al.}\cite{Gonzalez} have derived a
Kondo Hamiltonian of the type studied in Ref.\ \onlinecite{Romeike} from an
electronic model. They have shown that a
transverse magnetic field can induce or quench the Kondo effect. This is due to
Berry-phase interference between different quantum tunneling paths of the spin.

Koerting \textit{et al.}\cite{Koerting} consider the nonequilibrium Kondo
effect for a double quantum dot with four leads. By removing the leads
coupled to one of the dots one would obtain a model similar to ours. The main
difference is that we include charge fluctuations on the dot which is coupled to
the leads, whereas Koerting \textit{et al.}\cite{Koerting}
work in the regime of weak tunneling and Coulomb blockade, where both
quantum dots act as local spins.

In the present paper, we address the question of spin-dependent resonant
tunneling and Kondo tunneling through molecular magnets. As noted above, we
consider a resonant-tunneling model where the electron spin on the quantum
dot or molecule is coupled to an additional local, anisotropic spin via an
exchange interaction $J$. We assume this interaction to be weak, which
allows us to employ perturbation theory for small $J$. The
stationary current through the left and right leads is
identical and is related to the local electronic spectral function
$A_{\sigma}(\omega)$ on the molecule by the Meir-Wingreen formula\cite{Meir}
\begin{equation}\label{Eq1}
\langle I_L \rangle = \frac{e}{2\pi\hbar} \sum_{\sigma} \int d\omega \,
\frac{\Gamma_L\Gamma_R}{\Gamma_L+\Gamma_R} \left[f_L(\omega)-f_R(\omega) \right]
  \, A_{\sigma}(\omega),
\end{equation}
where $\sigma$ is the spin, $\Gamma_\alpha$ is the broadening of the
molecular level due to the hybridization with lead $\alpha=L,R$, to be defined
below, and $f_{\alpha}$ denotes the Fermi distribution function of lead
$\alpha$. The spectral function is determined by the imaginary part of the
retarded Green's function, $A_{\sigma}(\omega)=-2\,\text{Im}
G^\text{ret}_{\sigma\sigma}(\omega)$.
If transport is dominated by a single molecular level of energy $\varepsilon_d$,
the coupling to the leads gives rise to a Lorentzian form of the spectral
function, $A^0_{\sigma}(\omega) = \Gamma /[(\omega-\varepsilon_d)^2+\Gamma^2/4]$
with $\Gamma=\Gamma_L+\Gamma_R$, which manifests itself as a peak in the
differential conductance. For single-molecule devices, the excitation of
additional degrees of freedom due to the electronic tunneling is expected to
translate into additional characteristic features in the current.

We consider two complementary situations. The first is the case of arbitrary
gate and bias voltages but excluding the region where the Kondo contribution to
the current is large. Within a master-equation approach treating the local
exchange interaction perturbatively to second order, we calculate the transition
rates between local-spin states, showing that the spin can be driven out of
equilibrium even for strong molecule-lead hybridization. Signatures of inelastic
tunneling such as the fine-structure splitting of the
differential-conductance peaks persist in the regime where the
hybridization energy exceeds the thermal energy.

The second case concerns the regime of a large Kondo contribution to
the differential conductance, which only occurs for small bias voltages on 
the order of $|eV| \sim K_2(2S-1)$, as we shall see. Here, transport is studied
using a diagrammatic approach. We consider the case that the molecular
orbital is far from resonance so that the resonant-tunneling contributions
are negligible. In addition, this allows us to obtain analytical expressions.
We find that the magnetic anisotropy gives rise to the appearance of
two Kondo peaks in the differential conductance at finite bias voltages $\pm
V_c$. This intrinsically nonequilibrium Kondo effect is quite different from
the zero-bias peak studied by Romeike \textit{et al.},\cite{Romeike} which
relies on higher-order anisotropies absent from our model. In our
case, the magnetic anisotropy acts like a magnetic field in that it
gives rise to a splitting of the Kondo peak. Furthermore, we find a suppression
of the differential conductance with $1/\varepsilon_d^6$.

The paper is organized as follows. In Sec.~\ref{SectionModel}, we introduce our
model. Section~\ref{SectionMasterEquation} considers transport within a
master-equation approach, which allows us to study magnetic nonequilibrium
phenomena, whereas Sec.~\ref{SectionKondo} considers a diagrammatic approach,
which applies to the Kondo regime. In Sec.~\ref{SectionConclusion} we summarize
and discuss our results further. Some detailed calculations are relegated to
Appendices.

\section{Model}\label{SectionModel}

We consider a magnetic molecule coupled to two metallic leads.
Electronic tunneling through the junction is assumed
to involve a single orbital with energy $\varepsilon_d$ and spin
$\mathbf{s}$ that is coupled to a local spin $\textbf{S}$ via 
exchange interaction. The model is described by the Hamiltonian
\begin{equation}
H = H_0 + H_J + H_{\text{mag}},
\end{equation}
where
\begin{align}\label{resonant-tunneling}
H_0 ~=~& \varepsilon_d \sum_\sigma d^\dagger_\sigma d_\sigma
  + \sum_{\alpha \mathbf{k}\sigma} \epsilon_{\alpha\mathbf{k}}\,
  a^\dagger_{\alpha\mathbf{k}\sigma} a_{\alpha\mathbf{k}\sigma} \nonumber \\
& + \sum_{\alpha\mathbf{k}\sigma} \left( t_\alpha\,
  a^\dagger_{\alpha\mathbf{k}\sigma}
  d_\sigma + t^*_\alpha\, d^\dagger_\sigma a_{\alpha\mathbf{k}\sigma} \right)
\end{align}
is the resonant-tunneling Hamiltonian,
\begin{equation}
H_J = J\, \mathbf{s}\cdot\mathbf{S}
\end{equation}
with $\mathbf{s} \equiv \sum_{\sigma\sigma'} d^\dagger_\sigma
({\mbox{\boldmath$\sigma$}_{\sigma\sigma'}}/{2}) d_{\sigma'}$ is the exchange
interaction between the electrons in the molecular orbital and the local spin
$\mathbf{S}$, and
\begin{equation}
H_{\text{mag}} = -K_2(S^z)^2
\end{equation}
describes the easy-axis magnetic anisotropy of the local spin. We choose
the \textit{z} axis as the easy axis. Here, $d^\dagger_\sigma$ creates an
electron with spin $\sigma$ and energy $\varepsilon_d$ on the molecule, while
$a^\dagger_{\alpha\mathbf{k}\sigma}$ creates an electron with energy
$\epsilon_{\alpha\mathbf{k}}$, wave vector $\mathbf{k}$, and spin $\sigma$
in lead $\alpha$, which is considered a noninteracting electron gas.
The vector $\mbox{\boldmath$\sigma$}\equiv(\sigma_x,\sigma_y,\sigma_z)$ denotes
the Pauli matrices. In break junctions produced by electromigration, the
onsite energy $\varepsilon_d$ can be tuned by applying a gate
voltage.\cite{Park,Jo,Heersche,Grose}

\section{Master equation for the spin}\label{SectionMasterEquation}

The presence of strong coupling between the molecule and the leads prevents us
from treating the hybridization term in Eq.~(\ref{resonant-tunneling})
perturbatively. However, since the Hamiltonian becomes bilinear in the limit of
vanishing exchange coupling, $J=0$, our strategy is to diagonalize
$H_0+H_\text{mag}$ exactly while treating $H_J$ as a perturbation up to
second order. This approach allows us to study the nonequilibrium dynamics of
the molecular spin at finite bias voltages for arbitrary molecule-lead coupling
strengths, provided that Kondo correlations do not lead to a diverging
contribution from higher-order terms in the expansion.

We start by rewriting $H_0$ in terms of new ope\-ra\-tors,\cite{Mahan,Mitra}
\begin{equation}
H_0 = \sum_{\alpha\mathbf{k}\sigma} \epsilon_{\alpha\mathbf{k}}\,
  c^\dagger_{\alpha\mathbf{k}\sigma} c_{\alpha\mathbf{k}\sigma},
\label{H0c.1}
\end{equation}
where
\begin{eqnarray}
a_{\alpha\mathbf{k}\sigma} & = & \sum_{\alpha'\mathbf{k}'}
  \eta^{\alpha\mathbf{k}}_{\alpha'\mathbf{k}'} c_{\alpha'\mathbf{k}'\sigma},
\label{transformation1} \\
d_{\sigma} & = & \sum_{\alpha\mathbf{k}} \nu_{\alpha\mathbf{k}}
  c_{\alpha\mathbf{k}\sigma},
\label{transformation1a}
\end{eqnarray}
and
\begin{eqnarray}
\eta^{\alpha\mathbf{k}}_{\alpha'\mathbf{k}'} & = & \delta_{\alpha \alpha'}
  \delta_{\mathbf{k}\mathbf{k}'}- \frac{t_\alpha
  \nu_{\alpha'\mathbf{k}'}}
  {\epsilon_{\alpha\mathbf{k}}-\epsilon_{\alpha'\mathbf{k}'}+i\,\delta},
\label{transformation2}
\\
\nu_{\alpha\mathbf{k}} & = & \frac{t_\alpha}
  {\epsilon_{\alpha\mathbf{k}}-\varepsilon_d-\sum_{\alpha'\mathbf{k}'}
  \frac{t_{\alpha'}^2}
  {\epsilon_{\alpha\mathbf{k}}-\epsilon_{\alpha'\mathbf{k}'} -i\,\delta}}.
\label{transformation3}
\end{eqnarray}
For simplicity we assume real tunneling amplitudes $t_\alpha$.
In terms of the new operators, the exchange interaction assumes the form
\begin{equation}\label{H_J}
H_J = J \sum_{\alpha\alpha'\mathbf{k}\mathbf{k}'\sigma\sigma'} 
\nu_{\alpha\mathbf{k}}^* \nu_{\alpha'\mathbf{k}'}\,
c^{\dagger}_{\alpha\mathbf{k}\sigma}
\frac{\mbox{\boldmath$\sigma$}_{\sigma\sigma'}}{2} 
c_{\alpha'\mathbf{k}'\sigma'} \cdot \mathbf{S} .
\end{equation}

The time evolution of the density matrix $\rho$ of the full system is described
by the von Neumann equation, $\dot{\rho} = -(i/\hbar)[H,\rho]$. The degrees of
freedom of the local spin are described by the reduced density matrix
\begin{equation}
\rho_J = \text{Tr}_\text{el} \, \rho,
\end{equation}
which is obtained by tracing out all electronic degrees of freedom.
Assuming that the large electronic subsystem, which acts as a spin
reservoir, is weakly perturbed by the exchange coupling, we replace the full
density matrix by the direct product  $\rho \simeq \rho_J \otimes
\rho_\text{el}$. We need a further approximation for the electronic density
matrix $\rho_\text{el}$. We assume that different chemical potentials $\mu_L$
($\mu_R$) are imposed for the left (right) lead far from the junction. It
would thus be natural to assume Fermi distributions
$f_\alpha(\omega)=f(\omega-\mu_\alpha)$ for the physical electrons created by
$a_{\alpha\mathbf{k}\sigma}^\dagger$. However, we need to make a reasonable
assumption on the transformed \textit{c} fermions appearing in Eqs.\
(\ref{H0c.1}) and (\ref{H_J}). Since
$c^\dagger_{L\mathbf{k}\sigma}$ ($c^\dagger_{R\mathbf{k}\sigma}$) creates
an electron in a state with vanishing probability density far from the
junction in the right (left) lead, we assume the
occupation numbers of these states to be described by
$f_\alpha(\omega)$.

Making use of the Markov approximation that $\rho_J$ changes slowly on the time
scale of electronic relaxation, we obtain
\begin{equation}\label{iterated-rho}
\dot{\rho}_J(t) = - \frac{1}{\hbar^2} \int_{-\infty}^{t} dt' \,
\text{Tr}_\text{el} \, \left[ H_J(t) , \left[ H_J(t') , \rho_J(t) \otimes
\rho_\text{el} \right] \right].
\end{equation}
Here, operators $O(t)$ with an explicit time argument, including
$\rho_J(t)$, are in the interaction picture, $O(t) =
e^{i(H_0+H_\text{mag})t/\hbar}\,O\, e^{-i(H_0+H_\text{mag})t/\hbar}$.
Note that second-or\-der perturbation theory in the exchange coupling gives the
first non-vanishing correction to the conductance, since the expectation value
$\langle \textbf{S} \rangle$ and thus all first-order terms vanish exactly due
to symmetry.

We are interested in the stationary state. The stationary density matrix
$\rho_J$ has to be diagonal in the basis of eigenstates $|m\rangle$ of $S^z$,
since the full Hamiltonian $H$ is invariant under rotation about the
\textit{z}-axis in spin space. Inserting Eq.~(\ref{H_J}) into
Eq.~(\ref{iterated-rho}) we thus obtain a Pauli master equation, also
called rate equations, of the form
\begin{align}\label{rate-eq}
\dot{P}_{m} ~=~& P_{m+1} R_{m+1\rightarrow m} + P_{m-1} R_{m-1\rightarrow
m} \nonumber \\
& ~- P_{m}\left( R_{m\rightarrow m+1} + R_{m\rightarrow m-1} \right) ~=~0
\end{align}
for the occupation probabilities $P_m$ of spin states $|m\rangle$ in the
stationary state. The transition rates read 
\begin{eqnarray}
\lefteqn{ R_{m\to m\pm 1} = |\langle m\pm 1 | S^\pm | m \rangle|^2 \,
  \frac{J^2/4}{2\pi\hbar} } \nonumber \\
& & {}\times \sum_{\alpha\alpha'} \int d\omega \,
  |\tilde{\nu}_{\alpha}(\omega)|^2\,
  |\tilde{\nu}_{\alpha'}(\omega-[\pm 2m+1]K_2)|^2 \nonumber \\
& & {}\times \left[1-f_{\alpha}(\omega) \right] f_{\alpha'}(\omega-[\pm
  2m+1]K_2).
\end{eqnarray}
The spectral functions are given by
\begin{equation}
|\tilde{\nu}_{\alpha}(\omega)|^2 = \frac{\Gamma_\alpha}
  {(\omega-\varepsilon_d)^2+\Gamma^2/4}
\end{equation}
with $\Gamma \equiv \Gamma_L + \Gamma_R$ and $\Gamma_\alpha\equiv 2\pi
t_\alpha^2 D_\alpha$. The densities of states for the leads, $D_\alpha$, are
taken as constants. Compared to Eq.~(\ref{transformation3}) we have
approximated the self-energy part of $\nu_{\alpha\mathbf{k}}$ by a constant and
absorbed a factor $2\pi D_\alpha$.

At zero temperature, the integrals can be evaluated analytically. In the limit
of large bias voltages, the rates approach the constant value
\begin{eqnarray}\label{rateslimit}
R_{m\rightarrow m\pm 1} & = & \frac{\pi J^2 \, \Gamma_L \Gamma_R}
  {2\pi\hbar \, \Gamma \left[\Gamma^2+(\pm 2m + 1)^2K_2^2 \right]} \nonumber \\
& & {}\times  |\langle m \pm 1 | S^\pm | m \rangle|^2.
\end{eqnarray}
On the other hand, at zero bias only the rates involving the absorption of
energy are finite, whereas the emission rates vanish.

Solving Eq.~(\ref{rate-eq}) allows us to compute the differential conductance of
the molecular junction. The current operator of lead $\alpha$ reads
\begin{eqnarray}\label{EquationCurrentOperator}
\lefteqn{ I_\alpha = - i \frac{e}{\hbar} \sum_{\mathbf{k}\sigma} t_\alpha \left(
   a^\dagger_{\alpha\mathbf{k}\sigma} d_\sigma -d^\dagger_\sigma
a_{\alpha\mathbf{k}\sigma} \right) } \nonumber \\
& & = i\frac{e}{\hbar} \sum_{\mathbf{k}\sigma} \!
 \sum_{\alpha'\mathbf{k}' \alpha''\mathbf{k}''} \!\!
 \Big( t_\alpha \nu_{\alpha'\mathbf{k}'}^*
 \eta^{\alpha\mathbf{k}}_{\alpha''\mathbf{k}''}
 c^\dagger_{\alpha'\mathbf{k}'\sigma} c_{\alpha''\mathbf{k}''\sigma} -
  \text{h.c.}\Big) . \nonumber \\
& & {}
\label{currentI}
\end{eqnarray}
In order to compute the spin-dependent contribution to the expectation value
$\langle I_\alpha \rangle \equiv \text{Tr} \, I_\alpha \rho =
\text{Tr}\,I_\alpha(t)\rho(t)$ of the total current, we use the iterative
solution of the von Neumann equation,
\begin{equation}
\rho(t) = -\frac{1}{\hbar^2} \int_{-\infty}^t dt' \int_{-\infty}^{t'} dt''
\left[ H_J(t'), \left[ H_J(t''),\rho(t'') \right] \right].
\end{equation}
A term containing $\rho(-\infty)$ has dropped out here since it is linear
in $H_J$ and thus vanishes upon taking the trace. Making use of the Markov
approximation we find
\begin{align}\label{currentJ0}
\langle I_\alpha \rangle ^{(2)} ~=~& -\frac{1}{\hbar^2} \int_{-\infty}^t dt'
  \int_{-\infty}^{t'} dt'' \nonumber \\
& \times \text{Tr} \, \left[ \left[ I_\alpha(t), H_J(t') \right], H_J(t'')
\right] \rho(t)
\end{align}
for the second-order term. Carrying out the time integrals and evaluating the
spin sums as explained in Appendix \ref{AppCurrent}, we obtain
\begin{eqnarray}\label{currentJ}
\lefteqn{ \langle I_L \rangle^{(2)} = \frac{e}{2\pi \hbar} \, \frac{J^2}{4}
  \sum_{\alpha\alpha'\alpha''}
  \Big( \Gamma_L - \delta_{L\alpha} \left[ \Gamma_L + \Gamma_R \right] \Big) }
  \nonumber \\
& & {}\times \sum_m P_m\, \bigg\{
  \frac{1}{2}\, |\langle m-1 | S^- | m \rangle|^2
  I_{\alpha\alpha'\alpha''}([-2m+1]K_2) \nonumber \\
& & \quad {}+ \frac{1}{2}\, |\langle m+1 | S^+ | m \rangle|^2
  I_{\alpha\alpha'\alpha''}([2m+1]K_2) \nonumber \\
& & \quad {}+ |\langle m | S^z | m \rangle|^2
  I_{\alpha\alpha'\alpha''}(0)\bigg\}
\end{eqnarray}
with
\begin{eqnarray}\label{currentIaaa}
\lefteqn{ I_{\alpha\alpha'\alpha''}(E) \equiv \int d\omega \,
  |\tilde{\nu}_{\alpha''}(\omega)|^2 } \nonumber \\
& & {}\times \Big\{ |\tilde{\nu}_{\alpha}(\omega)|^2 \,
  |\tilde{\nu}_{\alpha'}(\omega-E)|^2 \,
  \left[1-f_{\alpha}(\omega)\right] f_{\alpha'}(\omega-E) \nonumber \\
& & \quad {}- |\tilde{\nu}_{\alpha}(\omega)|^2 \,
  |\tilde{\nu}_{\alpha'}(\omega+E)|^2 \,
  f_{\alpha}(\omega) \left[1-f_{\alpha'}(\omega+E)\right] \Big\}. \nonumber \\
& & {}
\end{eqnarray}
Equations (\ref{currentJ}) and (\ref{currentIaaa}) give the first non-vanishing
correction to the zero-order current $\langle I_L\rangle^{(0)}$,
which is obtained from the Meir-Wingreen formula
[Eq.~(\ref{Eq1})] by inserting the spectral function of the
unperturbed system, $A^0_\sigma(\omega) =
\Gamma/[(\varepsilon_d-\omega)^2+\Gamma^2/4]$.
Note that Eq.~(\ref{Eq1}) with $A_\sigma=A^0_\sigma$ is recovered by inserting
the equilibrium density matrix $\rho^{0}$ and the current operator from
Eq.~(\ref{EquationCurrentOperator}) into $\langle I_L\rangle^{0} = \text{Tr}\,
I_L\,\rho^{0}$.

A simple interpretation of Eq.~(\ref{currentJ}) is possible for the special case
of a 
local spin of length $S=1/2$, for which the magnetic anisotropy $K_2$ is
irrelevant and can be set to zero. For this case we obtain
\begin{align}
\langle I_L \rangle^{(2)} &~=~  \frac{e}{2\pi \hbar} \, \frac{J^2 S(S+1)}{4} \,
\frac{\Gamma_L\Gamma_R}{\Gamma} \nonumber \\
\times \int & d\omega \left[ \frac{\Gamma}
  {(\omega-\varepsilon_d)^2+\Gamma^2/4}\right]^3 \,
  \left[f_{L}(\omega) -f_{R}(\omega)\right].
\end{align}
Here, the third power of the spectral function appears, since the
current operator and the two exchange-interaction operators in
Eq.\ (\ref{currentJ0}) are each bilinear in fermionic operators.

\begin{figure}[!t]
\begin{center}
$\begin{array}{cc}
\textbf{(a)} \\ 
\includegraphics[height=5.5cm,angle=0]{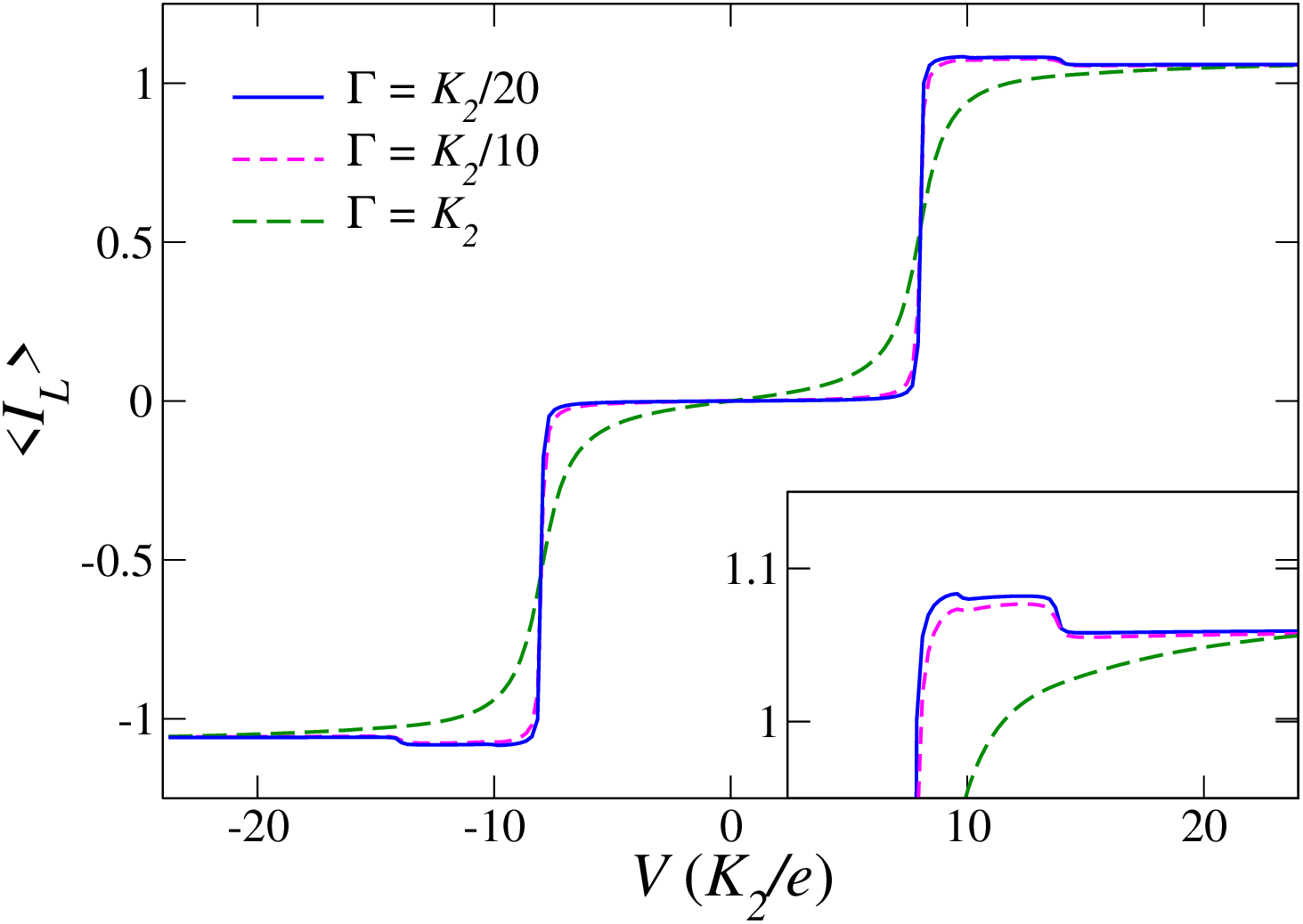} \\
\textbf{(b)} \\
\includegraphics[height=5.5cm,angle=0]{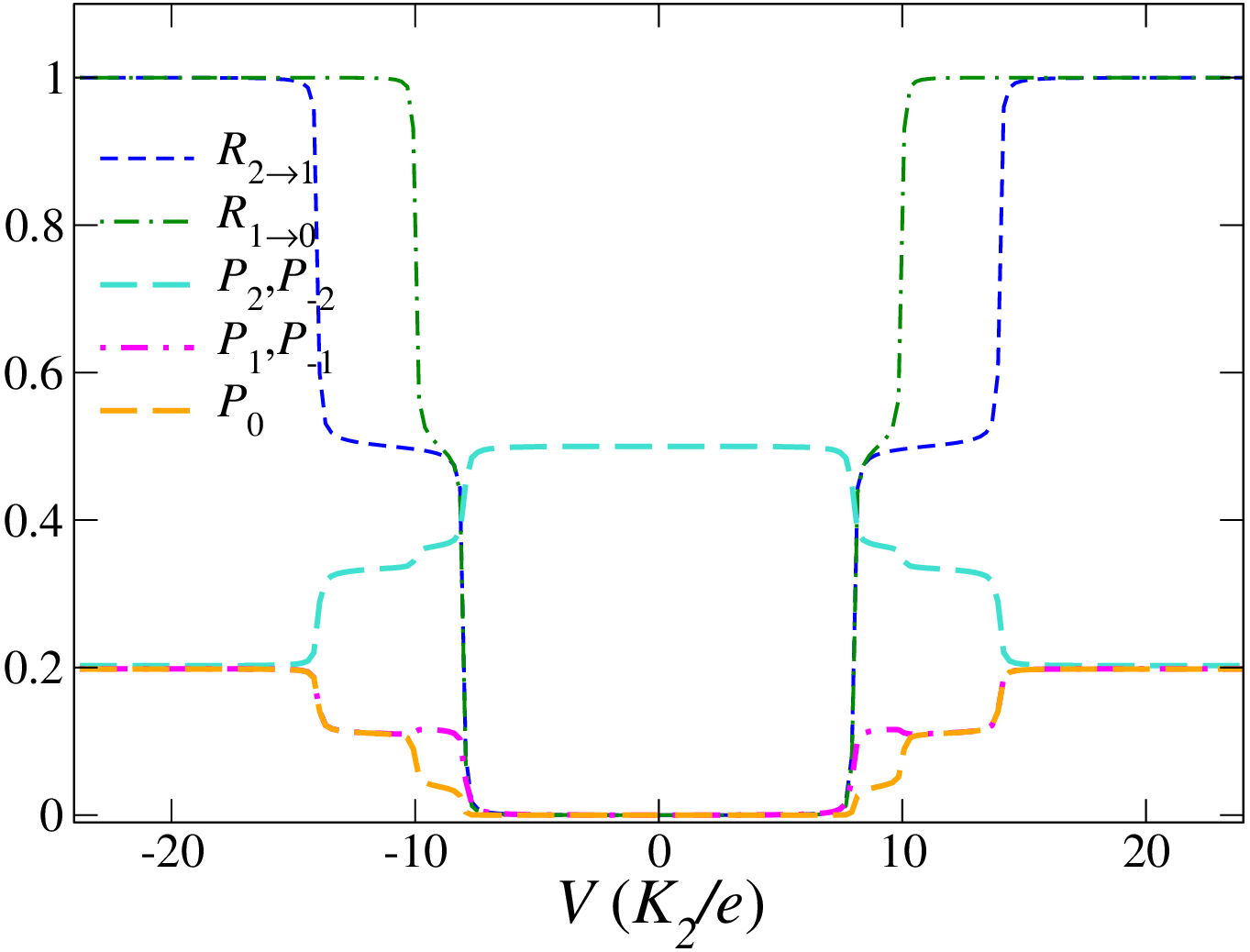}
\end{array}$
\caption{(Color online) (a) Current-voltage characteristics for different
hybridization
energies, $\Gamma=K_2/20$, $\Gamma=K_2/10$, and $\Gamma=K_2$. The inset
shows a closeup of the fine structure at positive bias. (b) Magnetic
transition rates $R_{2\rightarrow 1}$, $R_{1\rightarrow 0}$ and occupation
probabilities $P_m$ as functions of bias $V$ for $\Gamma=K_2/20$. We assume
symmetric couplings to the leads, $\Gamma_L=\Gamma_R$, and symmetric
capacitances, $\mu_L=eV/2$, $\mu_R=-eV/2$, a local molecular spin of length
$S=2$,
$\varepsilon_d=4 K_2$, and zero temperature. Further, we set $J=\Gamma/5$.
Currents are given in units of $(2e/\hbar) \Gamma_L \Gamma_R/\Gamma$.
Rates are given in units of their maximum values,
cf.~Eq.~(\ref{rateslimit}).}\label{FIG_current_symmetric}
\end{center}
\end{figure}

\begin{figure}[!t]
\begin{center}
$\begin{array}{cc}
\textbf{(a)} \\ 
\includegraphics[height=5.5cm,angle=0]{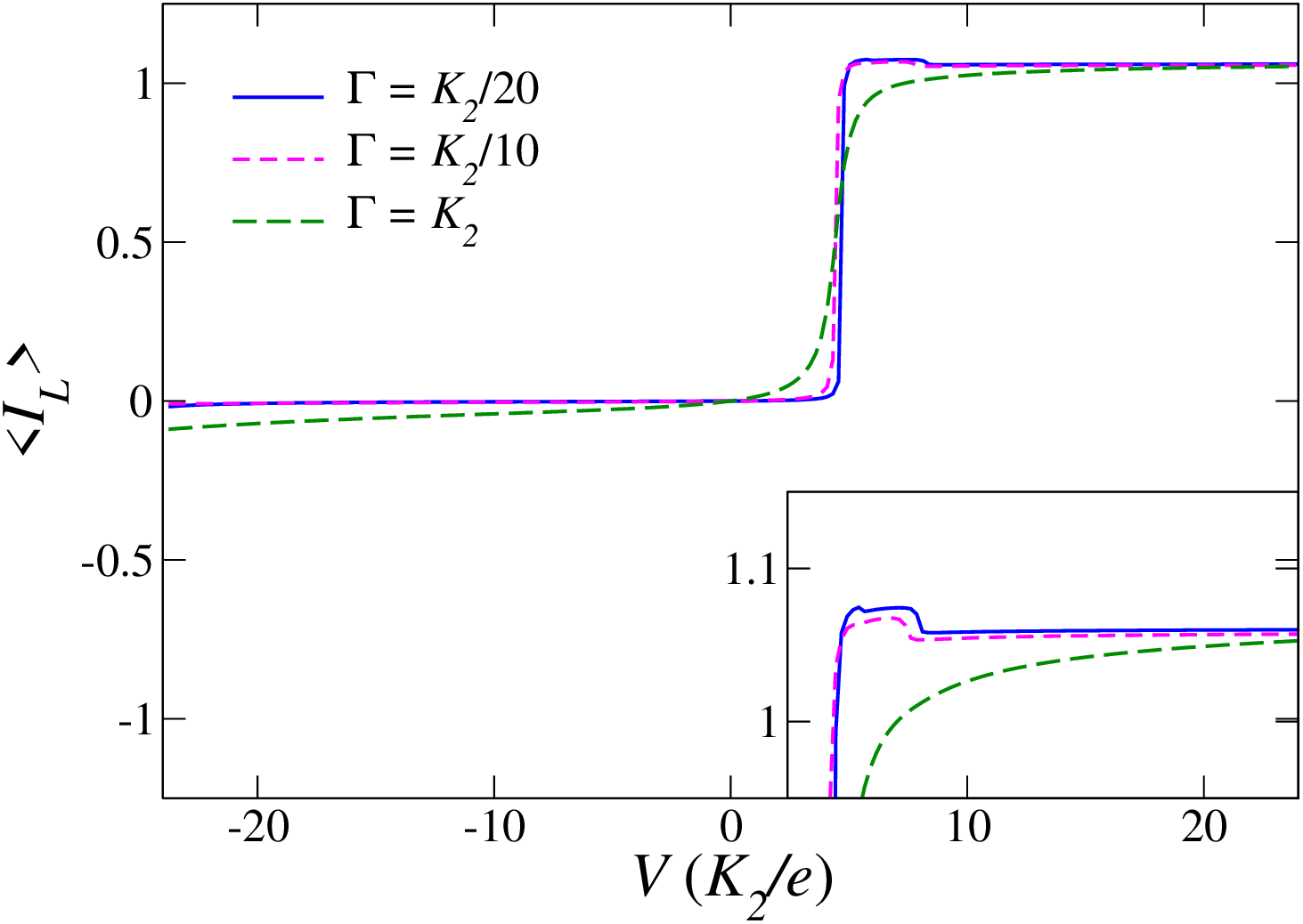} \\
\textbf{(b)} \\
\includegraphics[height=5.5cm,angle=0]{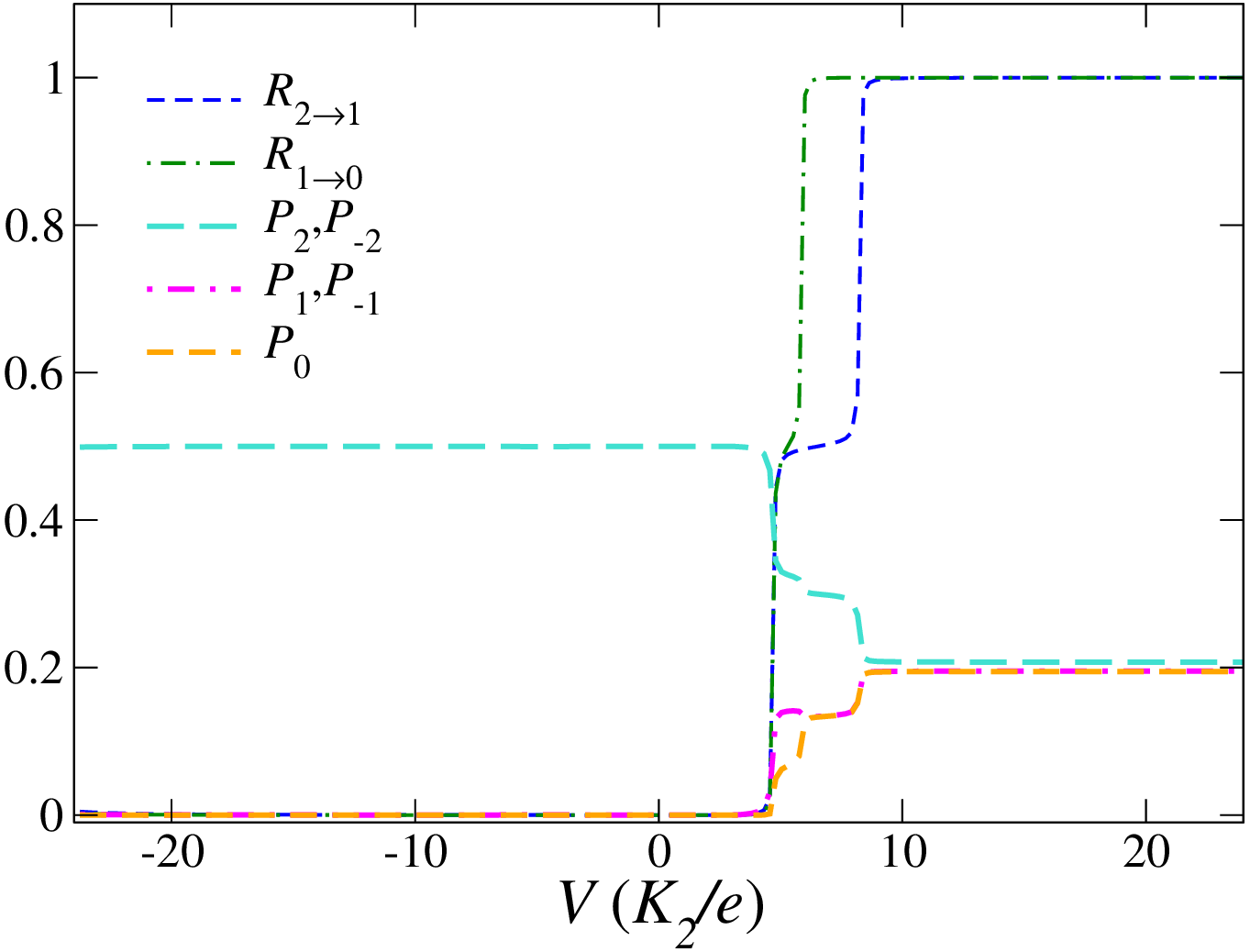}
\end{array}$
\caption{(Color online) (a) Current-voltage characteristics for different
hybridization
energies, $\Gamma=K_2/20$, $\Gamma=K_2/10$, and $\Gamma=K_2$. The inset
shows a closeup of the fine structure at positive bias. (b) Magnetic
transition rates $R_{2\rightarrow 1}$, $R_{1\rightarrow 0}$ and occupation
probabilities $P_m$ as functions of bias $V$ for $\Gamma=K_2/20$. We assume
strongly asymmetric couplings to the leads, $\Gamma_L\ll\Gamma_R$, and
strongly asymmetric
capacitances, $\mu_L=eV$, $\mu_R=0$, a local molecular spin of length $S=2$,
$\varepsilon_d=4 K_2$, and zero temperature. Further, we set $J=\Gamma/5$.
Currents are given in units of $(2e/\hbar) \Gamma_L
\Gamma_R/\Gamma$. Rates are given in units of their maximum values,
cf.~Eq.~(\ref{rateslimit}).}\label{FIG_current_asymmetric}
\end{center}
\end{figure}

If the magnetic anisotropy is large compared to the hybridization energy,
$K_2\gg \Gamma$, the general expression for the current in Eq.~(\ref{currentJ})
simplifies to
\begin{align}\label{refertothisEq}
\langle I_L \rangle^{(2)} =~& \frac{e}{2\pi \hbar} \, \frac{J^2}{4}
  \sum_{\alpha\alpha'\alpha''} \!\!
  \Big( \Gamma_L - \delta_{L\alpha} \left[ \Gamma_L + \Gamma_R \right] \Big) 
I_{\alpha\alpha'\alpha''}(0) \nonumber \\
&~\times \sum_m P_m |\langle m | S^z | m \rangle|^2,
\end{align}
since the integrals $I_{\alpha\alpha'\alpha''}([\pm 2m+1]K_2)$ are negligible
compared to $I_{\alpha\alpha'\alpha''}(0)$. Assuming symmetric
capacitances, one finds, in the limit of large bias voltages,
\begin{eqnarray}
\langle I_L \rangle^{(0)} & \to &
  \frac{2e}{\hbar} \, \frac{\Gamma_L \Gamma_R}{\Gamma}, \\
\langle I_L \rangle^{(2)} & \to &
  \frac{2e}{\hbar} \, \frac{\Gamma_L \Gamma_R}{\Gamma} \,
\frac{3S(S+1)}{4} \, \frac{J^2}{\Gamma^2},
\end{eqnarray}
for the zero-order and second-order contribution, respectively. Note
that the ferromagnetic or antiferromagnetic sign of $J$ does not affect the
results in the present approximation.

We first consider the situation of symmetric molecule-lead couplings and
capacitances, i.e., $|t_L| = |t_R|$, $\Gamma_L = \Gamma_R$, and $\mu_L = eV/2$,
$\mu_R = -eV/2$. Figure \ref{FIG_current_symmetric}(a) shows the
current-voltage characteristics up to second order in $J$ for the case of a
local spin of length $S=2$. The characteristic fine structure of the current
step at the Coulomb-blockade threshold is due to the second-order contribution,
$\langle I_L\rangle^{(2)}$, whereas the main step is mostly coming from
$\langle I_L\rangle^{(0)}$. The fine structure persists
as long as the hybridization energy $\Gamma$ remains small compared to the
magnetic anisotropy $K_2$. Note that the broadening of the steps is due to
$\Gamma>0$, and not to the temperature, for which we assume $T\ll \Gamma$.
For bias voltages below $|eV|=2\varepsilon_d$,
the current and all magnetic excitations are thermally suppressed.
However, as soon as the chemical potential of one lead aligns with the
resonance of the molecule, the current increases to
its maximum value. The current-induced magnetic transitions become
energetically accessible at the same time, as
shown in Fig.~\ref{FIG_current_symmetric}(b),
resulting in nonequilibrium probabilities $P_m$ of the different spin states.
In the limit of large bias voltages all spin states are equally occupied,
$P_m =1/(2S+1)$.

Interestingly, the presence of magnetic anisotropy leads to negative
differential conductance in the vicinity of $|eV| = 2\varepsilon_d$.
The underlying mechanism shall be explained briefly. According to
Eq.~(\ref{refertothisEq}), the magnetic states with maximum quantum numbers
$m=\pm S$
dominate the current since the current is proportional to the average $\sum_m
P_m |\langle m | S^z | m \rangle|^2$. Each decrease in $P_{\pm S}$ thus causes a
decrease in the current. Therefore, the spin-dependent contribution to the
current is large at low bias voltages, where $P_{\pm 2}=1/2$ and $\sum_{m} P_m
m^2 = 4$, and small at high bias voltages, where $P_{\pm 2}=1/5$ and $\sum_{m}
P_m m^2 = 2$.

\begin{figure}[!t]
\begin{center}
$\begin{array}{c}
\textbf{(a)} \\
\hspace{0.0cm}\includegraphics[height=4.2cm,angle=0]{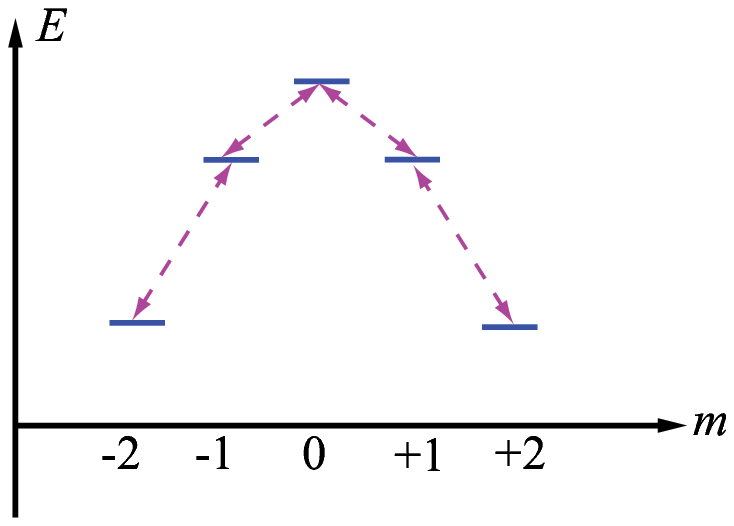} \\
\begin{array}{cc}
\textbf{(b)} & \hspace{0.2cm} \textbf{(c)} \\
\hspace{0.0cm}\includegraphics[height=4.2cm,angle=0]{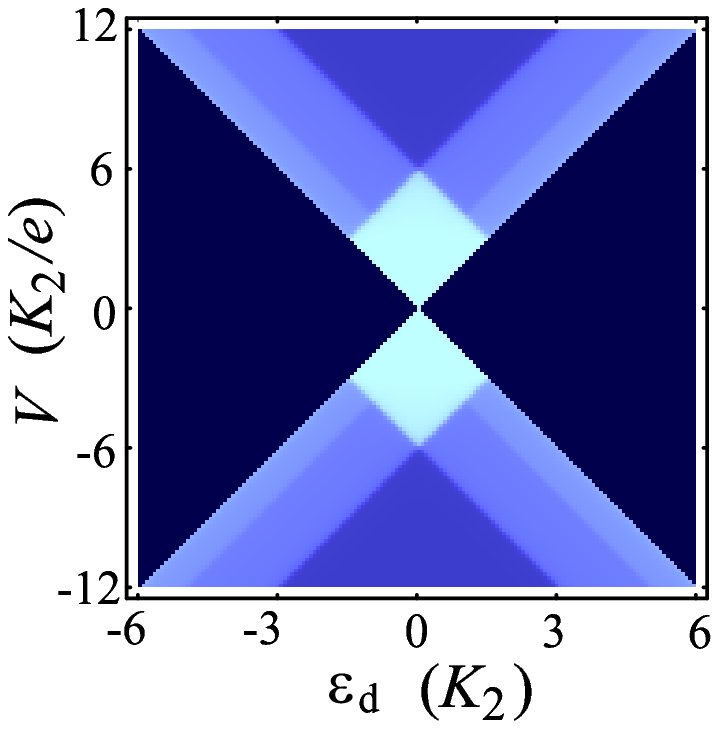} &
\hspace{0.0cm}\includegraphics[height=4.2cm,angle=0]{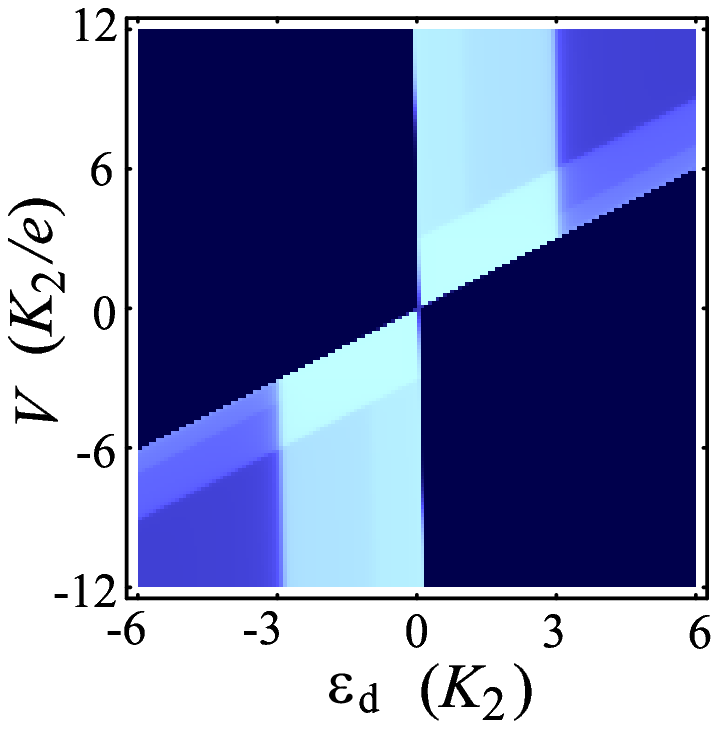}
\end{array}
\end{array}$
\caption{(Color online) (a) Level scheme showing
all spin transitions to order $J^2$. (b),(c) Two-dimensional density plots of
the absolute value of the second-order current contribution $\langle
I_L\rangle^{(2)}$ as a function of bias voltage $V$ and 
gate potential $\varepsilon_d$ for $\Gamma=K_2/20$. $\varepsilon_d$ is
controlled by the gate voltage.
We choose the same parameters as in Fig.~\ref{FIG_current_symmetric}.
Bright (dark) colors correspond to high (low) currents.
In (b) we assume symmetric couplings, $\mu_L=eV/2$, $\mu_R=-eV/2$,
$\Gamma_L=\Gamma_R$,
while in (c) we assume asymmetric couplings, $\mu_L=eV$, $\mu_R=0$, $\Gamma_R
\gg \Gamma_L$.}\label{FIG_2Dplots}
\end{center}
\end{figure}

We next consider the
situation of strongly asymmetric molecule-lead couplings and capacitances, i.e.,
$|t_L| \ll |t_R|$, $\Gamma_L \ll \Gamma_R$, and $\mu_L \simeq eV$, $\mu_R \simeq
0$. Note that this regime is naturally realized in many experimental setups,
whereas perfectly symmetric couplings are in general more difficult to achieve.
The current-voltage curves and the bias dependence of the magnetic excitation
rates are shown in Fig.~\ref{FIG_current_asymmetric}. Due to the asymmetric
coupling and $\varepsilon_d >0$, the current is suppressed for negative
bias voltages. However, the characteristic steps corresponding to excitations of
the molecular spin reappear at positive bias. Only their abscissas are reduced
by a factor
of 2, since the chemical potential of the left lead is now $\mu_L=eV$ instead
of $\mu_L=eV/2$. The device thus acts as a rectifier. Note the small ohmic
contribution with constant slope for large coupling $\Gamma$ in Fig.\
\ref{FIG_current_asymmetric}(a). We return to this point below.

We finally turn to the full bias and gate-voltage dependence of the
current. In both the symmetric and the asymmetric case,
selection rules for the spin require changes in the magnetic quantum number by
$\Delta m =0$ or $\Delta m = \pm 1$, where $\Delta m =0$ corresponds to elastic
and $\Delta m = \pm 1$ to inelastic scattering events, cf.\
Fig.~\ref{FIG_2Dplots}(a).
Inelastic scattering processes appear as additional steps in the current
and give rise to the magnetic fine structure in the two dimensional density
plots of the second-order contribution to the current as a function of bias and
gate voltages shown in Figs.~\ref{FIG_2Dplots}(b), (c).
We can now understand the origin of the weak ohmic conduction seen in Fig.\
\ref{FIG_current_asymmetric}(a) for $\Gamma=K_2$. What we are seeing is the
tail of the current step at $\epsilon_d=0$ in Fig.\ \ref{FIG_2Dplots}(c), which
is considerably broadened for $\Gamma=K_2$. Note that we here have
$\epsilon_d=4K_2=4\Gamma$, i.e., we are only $4\Gamma$ away
from the step. Since this distance does not depend on the bias voltage, the
conductivity is essentially constant, leading to ohmic behavior.

\section{Kondo transport}\label{SectionKondo}

Second-order perturbation theory in the exchange interaction $J$ fails, even
for small $J$, if the prefactors of higher-order terms diverge. This is the case
in the Kondo regime. Logarithmic divergences of the conductance first appear in
terms of \textit{third} order in
$J$.\cite{Ng,Glazman,Goldhaber-Gordon,Cronenwett,Wiel} (In this section, we
assume antiferromagnetic exchange, $J>0$.) Studying the emergence of
Kondo correlations thus requires to go beyond the second-order master-equation
approach discussed in Sec.~\ref{SectionMasterEquation}. For sufficiently small
$J$ and sufficiently large thermal energies, the conductance
is dominated by the third-order contribution, which we calculate in this
section. At lower temperatures, it would become necessary to resum the
divergences to all orders in
$J$.\cite{Ng,Glazman,Goldhaber-Gordon,Cronenwett,Wiel}

The total current through the molecule is related to the local electronic
spectral function $A_{\sigma}(\omega)=-2\text{Im}
G^\textit{ret}_{\sigma\sigma}(\omega)$
by the Meir-Wingreen formula, Eq.~(\ref{Eq1}),
where $G_{\sigma\sigma'}^\text{ret}(\omega) = \int d(t-t') \, e^{i\omega (t-t')}
G_{\sigma\sigma'}^\text{ret}(t-t')$
denotes the Fourier transform of the retarded single-particle Green's function
\begin{equation}
G^\text{ret}_{\sigma\sigma'}(t,t') \equiv -i \theta(t-t') \big\langle
  \big\{d_\sigma(t),d^\dagger_{\sigma'}(t')\big\} \big\rangle.
\end{equation}
Making use of the transformation defined in
Eqs.~(\ref{transformation1})--(\ref{transformation3}) requires
to compute the finite-temperature time-ordered Green's function
\begin{equation}\label{Greensfunction}
\mathcal{G}_{\alpha\alpha'\textbf{k}\textbf{k}'\sigma\sigma'}(\tau,\tau') 
  \equiv -\big\langle T_\tau \, c_{\alpha\textbf{k}\sigma}(\tau)
  c^\dagger_{\alpha'\textbf{k}'\sigma'}(\tau') \big\rangle.
\end{equation}
Our strategy is to expand
$\mathcal{G}_{\alpha\alpha'\textbf{k}\textbf{k}'\sigma\sigma'}$ 
in powers of $J$.

\begin{figure}[th]
\begin{center}
\includegraphics[height=4.0cm,angle=0]{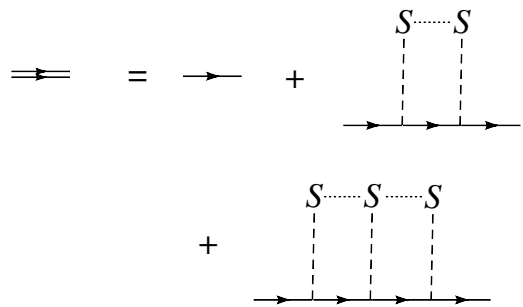}
\caption{Non-vanishing diagrams of the impurity Green's function up to third
order in $J$, following Ref.~\onlinecite{Bruus}. Diagrams including fermion
loops vanish exactly and are not shown. Spin averages are denoted by dotted
lines.}\label{diagrams}
\end{center}
\end{figure}

In order to obtain the current from the Meir-Wingreen formula, we need the
imaginary part of the electronic Green's function,
\begin{equation}
\text{Im} \sum_\sigma G^\text{ret}_{\sigma\sigma}(\omega) = \text{Im}
\sum_{\alpha\alpha'\textbf{k}\textbf{k}'\sigma} \nu_{\alpha\mathbf{k}}
\nu_{\alpha'\mathbf{k}'}^*
\mathcal{G}^\text{ret}_{\alpha\alpha'\textbf{k}\textbf{k}'\sigma\sigma}(\omega),
\label{Eq.ImG.2}
\end{equation}
where
$\mathcal{G}^\text{ret}_{\alpha\alpha'\textbf{k}\textbf{k}'\sigma\sigma}
(\omega)$ denotes the retarded Green's function.
All non-vanishing diagrams of the local Green's function up to third order
in $J$ are shown in Fig.~\ref{diagrams}, following the notation of
Ref.~\onlinecite{Bruus}. We again consider the case of strongly asymmetric
couplings, i.e., $|t_L| \ll
|t_R|$, $\Gamma_L \ll \Gamma_R$, and $\mu_L \simeq eV$, $\mu_R \simeq 0$.
As we shall see, Eq.~(\ref{Eq.ImG.2}) is then dominated by the contribution
from the right electrode, $\alpha=\alpha'=R$.
This allows us to describe the molecular degrees of freedom
by a thermal equilibrium distribution function that
is independent of the applied bias voltage and to obtain an
analytical expression.

The evaluation of Eq.~(\ref{Eq.ImG.2}) is shown in Appendix
\ref{AppGreensfunction}. We obtain
\begin{equation}
\text{Im} \sum_\sigma G^\text{ret}_{\sigma\sigma}(\omega)
\simeq -\frac{\Gamma/2}{\varepsilon_d^2+\Gamma^2/4} 
+ \left(\frac{\varepsilon_d}{\varepsilon_d^2+\Gamma^2/4}\right)^{\!2} \text{Im}
\Sigma^\text{ret}(\omega)
\end{equation}
with
\begin{widetext}
\begin{align}\label{selfenergy2nd3rd}
\text{Im} \, & \Sigma^\text{ret}(\omega) ~=~ - \frac{\pi}{2} J^2
\nu_0(\varepsilon_d) \sum_{mnl}
P_m\,
  \frac{1-f(\omega+E_m-E_l)}{1-f(\omega)}\,
  \Bigg\{ \delta_{nl} \sum_i |\langle m|S^i|n\rangle|^2
  \nonumber \\
& - i J \nu_0(\varepsilon_d) \sum_{ijk} \epsilon_{ijk} \langle
  m|S^i|n\rangle \langle n|S^j|l\rangle \langle l|S^k|m\rangle 
\bigg[ \ln \bigg| \frac{x}{\sqrt{(\omega+E_m-E_n)^2+T^2}}
  \bigg| + \ln \bigg| \frac{x}{\sqrt{(\omega+E_n-E_l)^2+T^2}}
  \bigg| \bigg] \Bigg\},
\end{align}
\end{widetext}
where
\begin{equation}
\nu_0(\varepsilon_d) = \frac{\Gamma/2\pi}{\varepsilon_d^2+\Gamma^2/4} .
\end{equation}
In the derivation we have assumed the molecular level to be far from
resonance, i.e., $|\varepsilon_d|$ to be large compared to $K_2S$, $T$, and
$\Gamma$, but still small compared to the band width $x$ of the leads. Details
are discussed in Appendix \ref{AppGreensfunction}. We have also assumed
$|\omega| \ll |\varepsilon_d|$, the significance of which will become clear in
the following step. Under these conditions, the resonant-tunneling
contribution to the differential conductance, which we have studied in Sec.\
\ref{SectionMasterEquation}, is negligible compared to the Kondo
contribution.

In the low-temperature limit $T\ll \Gamma$, derivatives of the Fermi
functions with respect
to the bias voltage become delta functions and the differential conductance
simplifies to
\begin{align}\label{Eqforfigure}
G \simeq \frac{e^2}{2\pi \hbar} & \,
\frac{\Gamma_L\Gamma_R}{\Gamma}\, \Bigg\{
\frac{\Gamma}{\varepsilon_d^2+\Gamma^2/4} \nonumber \\
& -2 \left(\frac{\varepsilon_d}{\varepsilon_d^2+\Gamma^2/4}\right)^2 \,
\text{Im}\, \Sigma^\text{ret}(eV) \Bigg\}.
\end{align}
Note that the argument $\omega$ of $\Sigma^\text{ret}(\omega)$ is $eV$. 
The assumption $|\omega| \ll |\varepsilon_d|$ made
above thus corresponds to $|\varepsilon_d|$ also being large compared to the
bias, $|eV|$. The spectral function has logarithmic divergences for $T\to 0$ at
the transition energies of the molecule, $E_m-E_n$, corresponding to virtual
transitions between two magnetic states $|m\rangle$ and $|n\rangle$. One
recovers the prefactor $3\pi J^2 D_0/8$, see Ref.~\onlinecite{Bruus},
for the case of spin $S=1/2$ and the (then irrelevant)
anisotropy set to $K_2=0$.

\begin{figure}[!t]
\begin{center}
\includegraphics[height=5.6cm,angle=0]{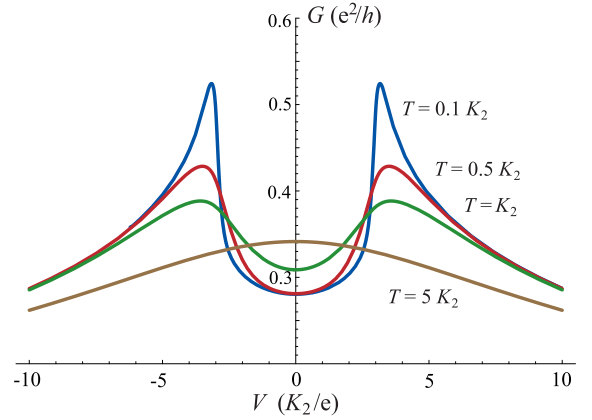}
\caption{(Color online) Differential conductance
$G$ in units of $e^2/h$ for different thermal energies $T$ (in units of
$K_2$) obtained from Eq.~(\ref{Eqforfigure}) as a function of bias
voltage $V$ in units of $K_2/e$. Here we assume a local molecular spin of length
$S=2$ and choose $J\nu_0(\varepsilon_d)=1$, $x=100\, K_2$,
and $\Gamma_R = 100 \, \Gamma_L$. Note that the parameters
$\Gamma$, $\varepsilon_d$, and $J$ leave
the curves for $G$ (in arbitrary units) invariant except for changing the
constant offset.}\label{FIG_kondo}
\end{center}
\end{figure}

Numerical results for nonzero temperatures are shown in Fig.~\ref{FIG_kondo}.
The
differential conductance $G$ diverges logarithmically for $T\to 0$ at critical
bias voltages $V=\pm V_c$ with
$eV_c=E_{S-1}-E_S = K_2(2S-1)$ since the emergence of Kondo correlations
requires the bias voltage to exceed the energy of the transition from the ground
states, $m=\pm S$, to the first excited states, $m=\pm(S-1)$.
Note that $G(V)$ is symmetric for positive and negative bias, in spite of
the highly asymmetric coupling since it is probing the electronic spectral
function. The situation is quite different from the case
considered by Romeike \textit{et al.},\cite{Romeike} which concerns a
zero-bias peak resulting from quantum tunneling between the two states
$|S\rangle$ and $|{-S}\rangle$. In our case, the splitting
of the Kondo peak as a consequence of magnetic anisotropy is more
similar to the situation of a quantum dot in an external magnetic field with
Zeeman energy $B$, where a Zeeman splitting of the energy levels leads to the
occurrence of two conductance peaks at $eV \simeq \pm B$ in the Kondo
regime.\cite{Paaske3}  At higher temperatures, $T\gg K_2$, the two
Kondo peaks merge into a single peak centered at zero bias due to the thermal
excitation of spin states with higher energy.

We now turn to the Kondo temperature $T_K$. Poor man's scaling for the
equilibrium case results in $T_K=0$ because the matrix elements of $S^\pm$
between the two degenerate ground states of the local spin vanish for our
model if $S>1/2$.\cite{Romeike}
Since a Kondo effect evidently does occur at nonzero bias, this
result is clearly not sufficient. A rough estimate of the Kondo temperature $T_K$ can be obtained as the
temperature for which the second-order and third-order terms
become equal in Eq.~(\ref{selfenergy2nd3rd}). We find 
$T_K \sim \exp[-1/\alpha \nu_0(\varepsilon_d) J]$, where $\alpha$
is a number of the order of unity. In the limit $K_2 \to 0$, where
the two peaks in Fig.~\ref{FIG_kondo} would merge, we recover the result $\alpha = 2$ for
an isotropic spin.

Since we focus on the case of strongly asymmetric couplings, where
the molecular degrees of freedom are in equilibrium with one of the two leads,
the logarithmic divergences are cut off by temperature 
or the applied bias voltage, respectively, in our perturbative approach, see
Eq.~(\ref{selfenergy2nd3rd}). The divergence for $T\to 0$ is
unphysical and would likely be removed by a resummation of higher-order
terms. By analogy to Ref.~\onlinecite{Paaske3}, we conjecture that the
divergence is ultimately cut off by a voltage-dependent
spin-relaxation rate.

While we have so far discussed the dependence on the bias voltage, see
Fig.\ \ref{FIG_kondo}, we now turn to the gate voltage. The gate voltage shifts
the on-site energy $\varepsilon_d$ and thus enters the expression for the
current through the square of the spectral function
$\Gamma/[\varepsilon_d^2+\Gamma^2/4]$
and the square of the factor $\varepsilon_d/[\varepsilon_d^2+\Gamma^2/4]$.
In particular, we obtain a suppression of $G\propto 1/\varepsilon_d^6$ in the
limit of strong detuning, $\varepsilon_d \gg \Gamma$.

\section{Conclusions}\label{SectionConclusion}

We have studied the spin-dependent electronic transport through magnetic
molecules for strong coupling to the leads.
Our discussion has focused on two complementary regimes.

For the first regime, we have presented
a description of transport in terms of a master equation
that keeps the electronic tunneling exactly, holds for arbitrary bias and
gate voltages, and treats the local exchange interaction $J$
perturbatively at second order. This approach is thus applicable for small $J$.
We have derived the bias-dependent magnetic transition rates showing that the
tunneling current can be used to drive the molecular spin out of equilibrium.
Further, we have shown that the characteristic fine structure
of the differential-conductance peaks persists for strong
molecule-lead coupling, where the broadening of the peaks is determined by the
hybridization energies.

The perturbative expansion in $J$ fails if Kondo correlations contribute
significantly to the transport. In this case, prefactors of the third- and
higher-order terms in $J$ diverge for $T\to 0$. The Kondo correlations can
become important for small bias voltages on the order of $|eV| \sim
K_2(2S-1)$. Here, transport is described by the
Meir-Wingreen formula in combination with a diagrammatic calculation of the
local electronic spectral function of the molecule. We have assumed the
molecular level to be far from resonance, which on the one hand makes sure that
the resonant-tunneling contributions to the conductance are small and which on
the
other allows us to obtain analytical results. We have shown that Kondo peaks
appear at finite bias voltages proportional to the anisotropy energy of the
molecular spin.

Our results leave several avenues for future research. First, it would be
interesting to include a local Coulomb interaction $U$ between the electrons on
the molecule.
However, due to the large hybridization there are no states with large
probability on the dot and the effect of $U$ is expected to be relatively weak.
We expect that for very large $U$
an equilibrium Kondo resonance could occur as a zero-bias peak in
the differential conductance in addition to the nonequilibrium Kondo effect
described in this paper.
Second, the presence of an external magnetic field might lead to an
interesting interplay with the splitting of the Kondo peaks due to the
magnetic anisotropy. Finally, it would be desirable to combine the two cases
studied here and to analyze the Kondo effect in magnetic molecules in the
resonant-tunneling regime, where resonant-tunneling contributions to the
conductance are not negligible and the spin is driven out of equilibrium by the
current.

\acknowledgments

We would like to thank A. Donabidowicz-Kolkowska, D.\ R.\ Reichman, and A.\
J.\ Millis for useful discussions. Financial support by the Deut\-sche
For\-schungs\-ge\-mein\-schaft is gratefully acknowledged.

\appendix

\section{Calculation of the current}\label{AppCurrent}

In this appendix we give details on the derivation of Eqs.~(\ref{currentJ})
and (\ref{currentIaaa}). We start from Eq.~(\ref{currentJ0}),
\begin{align}
\langle I_\alpha \rangle ^{(2)} ~=~& -\frac{1}{\hbar^2} \int_{-\infty}^t dt'
  \int_{-\infty}^{t'} dt'' \nonumber \\
& \times \text{Tr} \, \left[ \left[ I_\alpha(t), H_J(t') \right], H_J(t'')
\right] \rho(t).
\end{align}
Inserting the expressions for the current operator $I_\alpha$,
Eq.~(\ref{currentI}), and for the exchange interaction $H_J$, Eq.~(\ref{H_J}),
we find
\begin{eqnarray}
\lefteqn{ \langle I_L \rangle^{(2)} = i \frac{e}{\hbar} \frac{t_L J^2}{4\hbar^2}
\int_{-\infty}^t \!\!\! dt' \int_{-\infty}^{t'} \!\!\! dt'' \, \text{Tr} \,
\sum_{\textbf{k}\sigma} \sum_{123456}
\delta_{\sigma \sigma_1} \delta_{\sigma_1 \sigma_2} } \nonumber \\
& & {}\times \bigg( \eta^{L\textbf{k}*}_{1} \nu_{2} \nu_{3}^* \nu_{4}
\nu_{5}^* \nu_{6} - \nu_{1}^* \eta^{L\textbf{k}}_{2} \nu_{3}^* \nu_{4}
\nu_{5}^* \nu_{6} \bigg) \nonumber \\
& & {}\times \bigg[ \left[ c^\dagger_{1}(t) c_{2}(t), c^\dagger_{3}(t')
c_{4}(t') \, \mbox{\boldmath$\sigma$}_{\sigma_3\sigma_4}\cdot\mathbf{S}(t')
\right], \nonumber \\
& & \quad c^\dagger_{5}(t'') c_{6}(t'')
  \mbox{\boldmath$\sigma$}_{\sigma_5\sigma_6}\cdot\mathbf{S}(t'') \bigg]
\rho(t),
\hspace{8em}
\end{eqnarray}
where we have assumed $t_L$ to be real.
Here, the shorthand notation $j=1,2,3,4,5,6$ stands for
$(\alpha_j,\mathbf{k}_j,\sigma_j)$.

Introducing $\tau=t-t'$ and $\tau'=t'-t''$ and assuming a product state
gives
\begin{widetext}
\begin{eqnarray}
\lefteqn{ \langle I_L \rangle^{(2)} = i \frac{e}{\hbar} \frac{t_L
  J^2}{4\hbar^2} \int_{0}^\infty d\tau \int_{0}^\infty d\tau' \,
  \sum_{\textbf{k}} \sum_{123456} \big( \eta^{L\textbf{k}*}_{1} \nu_{2}
  \nu_{3}^* \nu_{4}
  \nu_{5}^* \nu_{6} - \nu_{1}^* \eta^{L\textbf{k}}_{2} \nu_{3}^* \nu_{4}
  \nu_{5}^* \nu_{6} \big) } \nonumber \\
& & {}\times \Big\{
  \big[ \delta_{23}\delta_{45}\delta_{16} f_2 f_4 \left( 1 - f_6 \right)
  e^{i(\epsilon_6-\epsilon_2)\tau/\hbar} e^{i(\epsilon_6-\epsilon_4)\tau'/\hbar}
  - \delta_{14}\delta_{25}\delta_{36} f_2 \left( 1 - f_4 \right) \left(
  1 - f_6 \right) e^{i(\epsilon_4-\epsilon_2)\tau/\hbar}
  e^{i(\epsilon_6-\epsilon_2)\tau'/\hbar} \nonumber \\
& & \qquad{}- \delta_{14}\delta_{25}\delta_{36} f_2 f_4 \left( 1 - f_6 \right)
  e^{i(\epsilon_4-\epsilon_2)\tau/\hbar} e^{i(\epsilon_6-\epsilon_2)\tau'/\hbar}
  + \delta_{23}\delta_{45}\delta_{16} \left( 1 - f_2 \right) f_4
  \left( 1  - f_6 \right) e^{i(\epsilon_6-\epsilon_2)\tau/\hbar}
  e^{i(\epsilon_6-\epsilon_4)\tau'/\hbar} \big] \nonumber \\
& & \qquad{}\times
  \text{Tr}_J \, 2\, \mathbf{S}(0)\cdot\mathbf{S}(-\tau') \,  \rho_J \nonumber
\\
& & \quad{}- \big[ \delta_{23}\delta_{45}\delta_{16} f_2 \left( 1 - f_4 \right)
  f_6
  e^{i(\epsilon_6-\epsilon_2)\tau/\hbar} e^{i(\epsilon_6-\epsilon_4)\tau'/\hbar}
  - \delta_{14}\delta_{25}\delta_{36} \left( 1 - f_2 \right) \left( 1 -
  f_4 \right) f_6 e^{i(\epsilon_4-\epsilon_2)\tau/\hbar}
  e^{i(\epsilon_6-\epsilon_2)\tau'/\hbar}  \nonumber \\
& & \qquad{}- \delta_{14}\delta_{25}\delta_{36} \left( 1 - f_2 \right) f_4 f_6
  e^{i(\epsilon_4-\epsilon_2)\tau/\hbar} e^{i(\epsilon_6-\epsilon_2)\tau'/\hbar}
  + \delta_{23}\delta_{45}\delta_{16} \left( 1 - f_2 \right) \left( 1 -
  f_4 \right) f_6 e^{i(\epsilon_6-\epsilon_2)\tau/\hbar}
  e^{i(\epsilon_6-\epsilon_4)\tau'/\hbar} \big] \nonumber \\
& & \qquad{}\times \text{Tr}_J \, 2\, \mathbf{S}(0)\cdot\mathbf{S}(\tau') \,
  \rho_J \Big\}.
\end{eqnarray}
This result can be rewritten as
\begin{eqnarray}
\lefteqn{ \langle I_L \rangle^{(2)} = i \frac{e}{\hbar} \frac{t_L
  J^2}{4\hbar^2} \int_{0}^\infty \!\! d\tau \int_{0}^{\infty} \!\! d\tau' \,
  \sum_{\textbf{k}} \sum_{123456} \big( \eta^{L\textbf{k}*}_{1} \nu_{2}
  \nu_{3}^* \nu_{4} \nu_{5}^* \nu_{6} - \nu_{1}^* \eta^{L\textbf{k}}_{2}
  \nu_{3}^* \nu_{4} \nu_{5}^* \nu_{6} \big) } \nonumber \\
& & \!\!{}\times \Big\{ \big[ \delta_{23}\delta_{45}\delta_{16} f_4 \left( 1 -
  f_6 \right) e^{i(\epsilon_6-\epsilon_2)\tau/\hbar}
  e^{i(\epsilon_6-\epsilon_4)\tau'/\hbar} - \delta_{14}\delta_{25}\delta_{36}
  f_2 \left( 1 - f_6 \right)
  e^{i(\epsilon_4-\epsilon_2)\tau/\hbar} e^{i(\epsilon_6-\epsilon_2)\tau'/\hbar}
  \big] 
  \text{Tr}_J \, 2\, \mathbf{S}(0)\cdot\mathbf{S}(-\tau') \, \rho_J
  \nonumber \\
& & \!\!{}- \big[ \delta_{23}\delta_{45}\delta_{16} \left( 1 - f_4 \right) f_6
  e^{i(\epsilon_6-\epsilon_2)\tau/\hbar} e^{i(\epsilon_6-\epsilon_4)\tau'/\hbar}
  - \delta_{14}\delta_{25}\delta_{36} \left( 1 - f_2 \right) f_6
  e^{i(\epsilon_4-\epsilon_2)\tau/\hbar} e^{i(\epsilon_6-\epsilon_2)\tau'/\hbar}
  \big] \text{Tr}_J \, 2\, \mathbf{S}(0)\cdot\mathbf{S}(\tau') \, \rho_J
  \Big\}. \nonumber \\
& & {}
\end{eqnarray}
\end{widetext}
The sums over spin indices are simplified by making use of the
identities
\begin{eqnarray}
\sum_{\sigma\sigma'}
\mbox{\boldmath$\sigma$}_{\sigma\sigma'}\cdot\mathbf{S}_1\,\mbox{
  \boldmath$\sigma$}_{\sigma'\sigma}\cdot\mathbf{S}_2 & = &
  2 \, \mathbf{S}_1 \cdot \mathbf{S}_2, \nonumber \\
\sum_{\sigma\sigma'}
\mbox{\boldmath$\sigma$}_{\sigma\sigma'}\cdot\mathbf{S}_1\,\mbox{
\boldmath$\sigma$}_{\sigma',-\sigma}\cdot\mathbf{S}_2 & = & 0.
\end{eqnarray}
In the coefficients $\nu_{\alpha\mathbf{k}}$ in Eq.~(\ref{transformation3}),
we approximate the self-energy part by a constant, as we did in
Sec.~\ref{SectionMasterEquation},
\begin{equation}
\nu_{\alpha\mathbf{k}} = \nu_{\alpha}(\epsilon_{\alpha\mathbf{k}}) \simeq 
\frac{t_\alpha}{\epsilon_{\alpha\mathbf{k}}-\varepsilon_d-i\Gamma/2}.
\end{equation}
Noting that Eqs.\ (\ref{transformation2}) and (\ref{transformation3}) imply
\begin{equation}
\sum_\textbf{k}\, \eta^{L\textbf{k}}_{1} = \delta_{L\alpha_1} + i\pi
D_L t_L \nu_{\alpha_1}(\epsilon_1)
\end{equation}
and
\begin{equation}
\nu_{\alpha}(\epsilon) - \nu_{\alpha}(\epsilon)^* = i \frac{\Gamma}{t_\alpha}
\, |\nu_{\alpha}(\epsilon)|^2,
\end{equation}
we arrive at the following expression for the tunneling current:
\begin{widetext}
\begin{eqnarray}
\lefteqn{ \langle I_L \rangle^{(2)} =
  i \frac{e}{\hbar} \frac{t_L J^2}{4\hbar^2} 
  \int_{0}^\infty \! d\tau \int_{0}^{\infty} \! d\tau' \sum_{123456}
  \big[ \left( \delta_{L\alpha_1} - i \pi D_L t_L \nu_{1}^* \right)
  \nu_{2} \nu_{3}^* \nu_{4} \nu_{5}^* \nu_{6}
  - \nu_{1}^* \left( \delta_{L\alpha_2} + i \pi D_L t_L \nu_{2} \right)
  \nu_{3}^* \nu_{4} \nu_{5}^* \nu_{6} \big] } \nonumber \\
& & \!\!{}\times \Big\{ \big[ \delta_{23}\delta_{45}\delta_{16} f_4 \left( 1 - f_6
  \right) e^{i(\epsilon_6-\epsilon_2)\tau/\hbar}
  e^{i(\epsilon_6-\epsilon_4)\tau'/\hbar}
  - \delta_{14}\delta_{25}\delta_{36} f_2 \left( 1 - f_6 \right)
  e^{i(\epsilon_4-\epsilon_2)\tau/\hbar} e^{i(\epsilon_6-\epsilon_2)\tau'/\hbar}
  \big] \text{Tr}_J \, 2\, \mathbf{S}(0)\cdot\mathbf{S}(-\tau') \, \rho_J
  \nonumber \\
& & \!\!{}-\big[ \delta_{23}\delta_{45}\delta_{16} \left( 1 - f_4 \right) f_6
  e^{i(\epsilon_6-\epsilon_2)\tau/\hbar} e^{i(\epsilon_6-\epsilon_4)\tau'/\hbar}
  - \delta_{14}\delta_{25}\delta_{36} \left( 1 - f_2 \right) f_6 
  e^{i(\epsilon_4-\epsilon_2)\tau/\hbar} e^{i(\epsilon_6-\epsilon_2)\tau'/\hbar}
  \big] \text{Tr}_J \, 2\, \mathbf{S}(0)\cdot\mathbf{S}(\tau') \, \rho_J
  \Big\}. \nonumber \\
\end{eqnarray}
\end{widetext}
Since we have assumed $\rho_J$ to be diagonal in the stationary state,
we finally obtain Eqs.~(\ref{currentJ}) and (\ref{currentIaaa}).

\section{Calculation of the impurity Green's function}\label{AppGreensfunction}

In order to use the Meir-Wingreen formula for the conductance, we have to compute
the imaginary part of the Green's function in Eq.~(\ref{Eq.ImG.2}). We consider
the situation of strongly asymmetric molecule-lead couplings and capacitances,
i.e., $|t_L| \ll |t_R|$, $\Gamma_L \ll \Gamma_R$, and $\mu_L \simeq eV$, $\mu_R
\simeq 0$.

Since Wick's theorem does not apply to spin operators, averages of products of
spin operators do not factorize into averages of pairs. We follow
Ref.~\onlinecite{Bruus} in evaluating the spin averages.
Expanding the electronic Matsubara-Green's function in powers of $J$ and
organizing the expansion in terms of
topologically distinct diagrams, one obtains\cite{Bruus}
\begin{widetext}
\begin{align}\label{appB.Green}
\mathcal{G}_{\alpha\alpha'\textbf{k}\textbf{k}'\sigma\sigma'}(\tau,\tau') & = 
\sum_{n=0}^{\infty} \left(-\frac{J}{\hbar}\right)^n \int_0^\beta d\tau_1 \cdots \int_0^\beta d\tau_n 
 \sum_{i_1 \cdots i_n}
\sum_{\sigma_1\cdots\sigma_n, \sigma'_1\cdots\sigma'_n}
 \Big\langle T_\tau \left[ S^{i_1}(\tau_1) \cdots S^{i_n}(\tau_n) \right]
\Big\rangle_0 \nonumber \\
& ~~\times \bigg\langle T_\tau \left[ B^{\dagger}_{\sigma_1}(\tau_1)
\frac{\mbox{\boldmath$\sigma$}_{\sigma_1\sigma'_1}}{2} B_{\sigma_1'}(\tau_1)
\cdots B^{\dagger}_{\sigma_n}(\tau_n)
\frac{\mbox{\boldmath$\sigma$}_{\sigma_n\sigma'_n}}{2} B_{\sigma_n'}(\tau_n)
c_{\alpha\textbf{k}\sigma}(\tau) c^\dagger_{\alpha'\textbf{k}'\sigma'}(\tau')
\right] \bigg\rangle_0,
\end{align}
\end{widetext}
where $\beta\equiv 1/T$ denotes the inverse thermal energy. For convenience, we
have defined $B_\sigma \equiv \sum_{\alpha \textbf{k}} \nu_{\alpha \textbf{k}}
c_{\alpha \textbf{k} \sigma}$. All non-vanishing diagrams up to third order in
$J$ are shown in Fig.~\ref{diagrams}. The linear term vanishes, since $\langle
\mathbf{S} \rangle = 0$. Diagrams with fermion loops are zero for the following
reasons:\cite{Bruus} a loop with a single fermion line results in taking the
trace of the Pauli matrix in the vertex, which yields zero. A loop with two
fermion lines appearing in the third-order diagrams gives rise to a trace over
two Pauli matrices, $\text{Tr}\, \sigma^i \sigma^j = 2\delta_{ij}$. The
resulting spin average $\langle T_\tau [S^{i_1}(\tau_1) S^{i_2}(\tau_2)
S^{i_3}(\tau_3)] \rangle_0$, with at least two of $i_1$, $i_2$, and $i_3$ equal,
vanishes.

Splitting off the zero-order term, the Green's function in
Eq.~(\ref{appB.Green}) can be written as\cite{Bruus}
\begin{eqnarray}
\lefteqn{ \mathcal{G}_{\alpha\alpha'\textbf{k}\textbf{k}'\sigma\sigma'}(\tau,
\tau') = \mathcal{G}^0_{\alpha\textbf{k}\sigma}(\tau,\tau')
\delta_{\alpha\alpha'} \delta_{\textbf{k}\textbf{k}'} \delta_{\sigma\sigma'} }
\nonumber \\
& & {}+ \int_0^\beta d\tau_1 \int_0^\beta d\tau_2 \,
\mathcal{G}^0_{\alpha\textbf{k}\sigma}(\tau,\tau_1)
\Sigma_{\alpha\alpha'\textbf{k}\textbf{k}'\sigma\sigma'}(\tau_1,\tau_2)
\nonumber \\
& & \quad{}\times \mathcal{G}^0_{\alpha'\textbf{k}'\sigma'}(\tau_2,\tau'),
\label{appB.Green.4}
\end{eqnarray}
where the unperturbed Matsubara-Green's function in the imaginary-time domain
is given by
\begin{equation}\label{appB.Green.5}
\mathcal{G}^0_{\alpha\textbf{k}\sigma}(\tau,\tau') = - \left[ \theta(\tau-\tau')
  - f(\omega_{\alpha\textbf{k}}) \right]
  e^{-\omega_{\alpha\textbf{k}}(\tau-\tau')/\hbar}
\end{equation}
with $\omega_{\alpha\textbf{k}} \equiv
\epsilon_{\alpha\mathbf{k}}-\mu_\alpha$. In the frequency domain we have
\begin{equation}
\mathcal{G}^0_{\alpha\textbf{k}\sigma}(i\omega_n) =
\frac{1}{i\omega_n-\omega_{\alpha\textbf{k}}} ,
\end{equation}
where $i\omega_n$ is a fermionic Matsubara frequency. Note that we are only
interested in the spin trace of the self-energy, $\sum_\sigma
\Sigma_{\alpha\alpha'\textbf{k}\textbf{k}'\sigma\sigma}^\text{ret}$, which
enters in
the Meir-Wingreen formula.

The second-order term of the self-energy yields
\begin{eqnarray}\label{appB.Sigma2.2}
\lefteqn{ \sum_\sigma
  \Sigma_{\alpha\alpha'\textbf{k}\textbf{k}'\sigma\sigma}^{\text{(2)}}
  (\tau_1,\tau_2) =
\frac{J^2}{2\hbar^2} \sum_{\alpha_1\mathbf{k}_1\sigma_1} \sum_{m,n}
\mathcal{G}^0_{\alpha_1\textbf{k}_1\sigma_1}(\tau_1,\tau_2) } \nonumber \\
& & {}\times \nu_{\alpha\mathbf{k}}^* \nu_{\alpha'\mathbf{k}'}
  |\nu_{\alpha_1\mathbf{k}_1}|^2 \nonumber \\
& & {}\times \sum_i
  |\langle m|S^i|n\rangle|^2 e^{(E_m-E_n)(\tau_1-\tau_2)/\hbar} P_m,
  \hspace{4em}
\end{eqnarray}
where we have used that $\text{Tr}\, \sigma^i\sigma^j=2\delta_{ij}$. Here,
$E_m\equiv -K_2m^2$ denotes the magnetic anisotropy energy in the spin state
$|m\rangle$ with occupation probability $P_m$. The spin averages in
Eq.~(\ref{appB.Green}) are to be evaluated for the unperturbed Hamiltonian
$H_\mathrm{mag}$,\cite{Bruus} leading to $P_m\propto e^{-\beta E_m}$. We
restrict ourselves to the off-resonance situation, i.e., the dark region in
Fig.~\ref{FIG_2Dplots}(c) with negligible resonant-tunneling differential
conductance, where the spin essentially remains in equilibrium. This is
certainly satisfied if
$|\varepsilon_d|$ is large compared to the energy scales relevant for the
Kondo contributions, $K_2S$ and $T$.

Equation (\ref{appB.Sigma2.2}) contains a sum over leads, $\alpha_1=L,R$,
and a factor of $t_{\alpha_1}^2$ under the sum. Since we have assumed strongly
asymmetric couplings, $|t_L| \ll |t_R|$, the sum is dominated by the
contribution from the right lead, $\alpha_1=R$. Dropping the term with
$\alpha_1=L$, we note that the Green's function
$\mathcal{G}^0_{R\mathbf{k}\sigma}(\tau,\tau')$ in Eq.~(\ref{appB.Green.5}) only
contains the Fermi distribution function for the right lead, which is
$f_R(\epsilon_{R\mathbf{k}})=f(\epsilon_{R\mathbf{k}})=
1/(e^{\beta\epsilon_{R\mathbf{k}}}+1)$, since $\mu_R=0$. Importantly, the
resulting expression is independent of the bias voltage.

Furthermore, we see that Eq.~(\ref{appB.Sigma2.2}) contains a factor
$t_\alpha t_{\alpha'}$. From Eq.~(\ref{Eq.ImG.2}) we obtain the same factor 
so that the contribution from leads $\alpha$, $\alpha'$ is
proportional to $t_\alpha^2 t_{\alpha'}^2$.
Since we have assumed $|t_L| \ll
|t_R|$, we can neglect all contributions except for $\alpha=\alpha'=R$. We will
keep only these contributions from now on.

Taking the Fourier transform of Eq.~(\ref{appB.Green.4}) and performing the
analytic continuation, we obtain the retarded Green's function
\begin{eqnarray}
\lefteqn{ G^\text{ret}_{RR\mathbf{k}\mathbf{k}'\sigma\sigma}(\omega)
= G^{\text{ret},0}_{RR\mathbf{k}\mathbf{k}'\sigma\sigma}(\omega) } \nonumber \\
& & {}+ \left[P\,\frac{1}{\omega-\epsilon_{R\mathbf{k}}}
  - i\pi\,\delta(\omega-\epsilon_{R\mathbf{k}})\right] \,
  \Sigma^\text{ret}_{RR\mathbf{k}\mathbf{k}'\sigma\sigma}(\omega) \nonumber \\
& & \quad{}\times \left[P\,\frac{1}{\omega-\epsilon_{R\mathbf{k}'}}
  - i\pi\,\delta(\omega-\epsilon_{R\mathbf{k}'})\right] ,
\label{Eq.Gret.omega.3}
\end{eqnarray}
where $P$ denotes the principal value. We assume $|\varepsilon_d|$ to be large
not only compared to $K_2S$ and $T$ but also to $\Gamma$. One can then
show that the delta-function terms are negligible compared to the principal
value terms. Including the factors of $\nu_{R\mathbf{k}}^\ast$,
$\nu_{R\mathbf{k}}$, we obtain expressions of the form
\begin{eqnarray}
\sum_{\mathbf{k}} |\nu_{R\mathbf{k}}|^2
  P\,\frac{1}{\omega-\epsilon_{R\mathbf{k}}}
& \simeq & -\frac{\Gamma_R}{\Gamma}\, \frac{\varepsilon_d -
  \omega}{(\varepsilon_d-\omega)^2
  + \Gamma^2/4} \nonumber \\
& \simeq & - \frac{\varepsilon_d - \omega}{(\varepsilon_d-\omega)^2
  + \Gamma^2/4} .
\end{eqnarray}
For the imaginary part of the Green's function in Eq.\ (\ref{Eq.ImG.2})
we then only require the imaginary part of $\sum_\sigma
\Sigma_{RR\textbf{k}\textbf{k}'\sigma\sigma}^{\text{ret},\text{(2)}}$ in Eq.\
(\ref{Eq.Gret.omega.3}). Taking the imaginary part of the Fourier transform of
Eq.~(\ref{appB.Sigma2.2}) we obtain
\begin{align}\label{self-energy2}
\text{Im} \, \sum_\sigma &
\Sigma_{RR\textbf{k}\textbf{k}'\sigma\sigma}^{\text{ret},\text{(2)}}(\omega) 
= - \frac{\pi J^2 D_0}{2} \sum_{m,n} \sum_i |\langle m|S^i|n\rangle|^2 \, P_m
\nonumber \\
\times \nu_{R\textbf{k}}^* & \nu_{R\textbf{k}'} |\nu_R(\omega+E_m-E_n)|^2
\frac{1-f(\omega+E_m-E_n)}{1-f(\omega)},
\end{align}
where we assume constant densities of states
for the leads, $D_0\equiv D_L=D_R$, and an energy band ranging
from $-x$ to $x$, where $x$ is the largest energy scale in our model.

The third-order term gives
\begin{align}
\sum_\sigma &
\Sigma_{RR\textbf{k}\textbf{k}'\sigma\sigma}^{\text{(3)}}
(\tau_1,\tau_2) = -\frac{J^3}{\hbar^3} \sum_\sigma \sum_{\textbf{k}_1 \sigma_1,
\textbf{k}_2 \sigma_2} \nu_{R\textbf{k}}^* \nu_{R\textbf{k}'} \nonumber \\
& \times |\nu_{R\textbf{k}_1}|^2 |\nu_{R\textbf{k}_2}|^2 \int_0^\beta d\tau_3\,
\mathcal{G}^0_{R\textbf{k}_1\sigma_1}(\tau_1,\tau_3)\,
\mathcal{G}^0_{R\textbf{k}_2\sigma_2}(\tau_3,\tau_2) \nonumber \\
& \times \sum_{ijk} \Big\langle T_\tau \left[ S^{i}(\tau_1) S^{j}(\tau_3)
S^{k}(\tau_2) \right] \Big\rangle_0
\frac{\sigma_{\sigma\sigma_1}^i}{2}
\frac{\sigma_{\sigma_1\sigma_2}^j}{2}
\frac{\sigma_{\sigma_2\sigma}^k}{2}.
\end{align}
Here, the average involving spin operators depends on the time arguments
$\tau_1$, $\tau_2$ and $\tau_3$, since
$i$, $j$ and $k$ can be different. However, since the self-energy only depends 
on the differences $\tau_1-\tau_2$ and $\tau_3-\tau_1$, we may set $\tau_2=0$
and distinguish the two possibilities
$\tau_1>\tau_3$ and $\tau_3>\tau_1$.
Using that $\text{Tr}[\sigma^i\sigma^j\sigma^k] = 2i \, \epsilon_{ijk}$,
inserting 
$\mathcal{G}^0_{R\textbf{k}\sigma}(\tau_1,\tau_3) = - \left[
\theta(\tau_1-\tau_3) - f(\epsilon_{R\textbf{k}}) \right]
e^{-\epsilon_{R\textbf{k}}(\tau_1-\tau_3)}$
and
$\mathcal{G}^0_{R\textbf{k}\sigma}(\tau_3,0) = - \left[ \theta(\tau_3) -
f(\epsilon_{R\textbf{k}}) \right] e^{-\epsilon_{R\textbf{k}}\tau_3}$,
and evaluating the integral over $\tau_3$, we obtain for $0\le \tau_1\le \beta$
\begin{widetext}
\begin{eqnarray}\label{timeintegral}
\sum_\sigma
  \Sigma_{RR\textbf{k}\textbf{k}'\sigma\sigma}^{\text{(3)}}(\tau_1,0) & = & -
  \frac{iJ^3}{4\hbar^2} \sum_{\mathbf{k}_1\mathbf{k}_2} \nu_{R\textbf{k}}^*
  \nu_{R\textbf{k}'} |\nu_{R\textbf{k}_1}|^2 |\nu_{R\textbf{k}_2}|^2
  \sum_{ijk} \epsilon_{ijk} \sum_{mnl} \langle m|S^i|n\rangle \langle
  n|S^j|l\rangle \langle l|S^k|m\rangle P_m \nonumber \\
& & {}\times \bigg\{
\frac{1-f(\epsilon_{R\mathbf{k}_1})}{\epsilon_{R\mathbf{k}_1}-\epsilon_{R\mathbf
{k}_2}+E_n-E_l} \left[1-f(\epsilon_{R\mathbf{k}_2})\right]
e^{-\epsilon_{R\mathbf{k}_2}\tau_1/\hbar} e^{(E_m-E_l)\tau_1/\hbar} \nonumber \\
& & \quad{}- \frac{1-f(\epsilon_{R\mathbf{k}_2})}{\epsilon_{R\mathbf{k}_1}-\epsilon_{R\mathbf
{k}_2}+E_n-E_l} \left[1-f(\epsilon_{R\mathbf{k}_1})\right]
e^{-\epsilon_{R\mathbf{k}_1}\tau_1/\hbar} e^{(E_m-E_n)\tau_1/\hbar} \nonumber \\
& & \quad{}- \frac{f(\epsilon_{R\mathbf{k}_1})}
  {\epsilon_{R\mathbf{k}_1}-\epsilon_{R\mathbf{k}_2}+E_m-E_n} 
  \left[1-f(\epsilon_{R\mathbf{k}_2})\right] e^{-\epsilon_{R\mathbf{k}_2}\tau_1/\hbar}
  e^{(E_m-E_l)\tau_1/\hbar} \nonumber \\
& & \quad{}+ \frac{f(\epsilon_{R\mathbf{k}_2})e^{\beta(E_m-E_n)}}{\epsilon_{R\mathbf{k}_1}
-\epsilon_{R\mathbf{k}_2}+E_m-E_n} 
\left[1-f(\epsilon_{R\mathbf{k}_1})\right] e^{-\epsilon_{R\mathbf{k}_1}\tau_1/\hbar}
e^{(E_n-E_l)\tau_1/\hbar} \bigg\} .
\end{eqnarray}
With $\langle m|S^i|n\rangle^* \langle n|S^j|l\rangle^* \langle
l|S^k|m\rangle^* = - \langle m|S^i|n\rangle \langle n|S^j|l\rangle \langle
l|S^k|m\rangle$ under the sum over $i$, $j$, $k$, Eq.~(\ref{timeintegral}) simplifies to
\begin{eqnarray}
\sum_\sigma \Sigma_{RR\textbf{k}\textbf{k}'\sigma\sigma}^{\text{(3)}}(\tau_1,0)
& = & - \frac{iJ^3}{2\hbar^2} \sum_{\mathbf{k}_1\mathbf{k}_2} \nu_{R\mathbf{k}}^*
  \nu_{R\mathbf{k}'} |\nu_{R\mathbf{k}_1}|^2 |\nu_{R\mathbf{k}_2}|^2
  \sum_{ijk} \epsilon_{ijk} \sum_{mnl} \langle m|S^i|n\rangle \langle
  n|S^j|l\rangle \langle l|S^k|m\rangle P_m \nonumber \\
& & {}\times \bigg\{
  \frac{1-f(\epsilon_{R\mathbf{k}_1})}{\epsilon_{R\mathbf{k}_1}-\epsilon_{R\mathbf
  {k}_2}+E_n-E_l} \left[1-f(\epsilon_{R\mathbf{k}_2})\right]
  e^{-\epsilon_{R\mathbf{k}_2}\tau_1/\hbar} e^{(E_m-E_l)\tau_1/\hbar} \nonumber \\
& & \quad{}- \frac{f(\epsilon_{R\mathbf{k}_1})}
  {\epsilon_{R\mathbf{k}_1}-\epsilon_{R\mathbf{k}_2}+E_m-E_n}  
  \left[1-f(\epsilon_{R\mathbf{k}_2})\right] e^{-\epsilon_{R\mathbf{k}_2}\tau_1/\hbar}
  e^{(E_m-E_l)\tau_1/\hbar} \bigg\} .
\end{eqnarray}
Computing the Fourier transform yields
\begin{eqnarray}
\sum_\sigma
\Sigma_{RR\textbf{k}\textbf{k}'\sigma\sigma}^{\text{ret},\text{(3)}}(\omega) & =
& - \frac{iJ^3}{2} \, \sum_{ijk} \epsilon_{ijk} \sum_{mnl} \langle
  m|S^i|n\rangle\langle n|S^j|l\rangle \langle l|S^k|m\rangle P_m 
  \sum_{\mathbf{k}_1\mathbf{k}_2} \nu_{R\mathbf{k}}^* \nu_{R\mathbf{k}'}
  |\nu_{R\mathbf{k}_1}|^2 |\nu_{R\mathbf{k}_2}|^2 \nonumber \\
& & {}\times \bigg\{
  \frac{1-f(\epsilon_{R\mathbf{k}_1})}{\epsilon_{R\mathbf{k}_1}-
    \epsilon_{R\mathbf  {k}_2}+E_n-E_l} \,
    \frac{1}{\omega-\epsilon_{R\mathbf{k}_2}+E_m-E_l+i\delta} \,
  \frac{1-f(\epsilon_{R\mathbf{k}_2})}{1-f(\epsilon_{R\mathbf{k}_2}-E_m+E_l)}
  \nonumber \\
& & \quad{}- \frac{f(\epsilon_{R\mathbf{k}_1})}
  {\epsilon_{R\mathbf{k}_1}-\epsilon_{R\mathbf{k}_2}+E_m-E_n}
  \, \frac{1}{\omega-\epsilon_{R\mathbf{k}_2}+E_m-E_l+i\delta} \,
  \frac{1-f(\epsilon_{R\mathbf{k}_2})}{1-f(\epsilon_{R\mathbf{k}_2}-E_m+E_l)}
  \bigg\}.
\label{appB.Sigma3.6}
\end{eqnarray}
\end{widetext}
The sum over $\mathbf{k}_2$ can be evaluated to give
\begin{widetext}
\begin{align}
\text{Im} \, \nu_{R\mathbf{k}} \nu_{R\mathbf{k}'}^* \, \sum_\sigma
\Sigma_{RR\textbf{k}\textbf{k}'\sigma\sigma}^{\text{ret},\text{(3)}}(\omega) &
~=~ \frac{i\pi D_0 J^3}{2} \, \sum_{ijk} \epsilon_{ijk} \sum_{mnl} \langle
  m|S^i|n\rangle\langle n|S^j|l\rangle \langle l|S^k|m\rangle P_m 
  \sum_{\mathbf{k}_1} |\nu_{R\mathbf{k}}|^2 |\nu_{R\mathbf{k}'}|^2
  |\nu_{R\mathbf{k}_1}|^2 \nonumber \\
\times |\nu_{R}(\omega+E_m-E_l)|^2 \bigg\{ &
\frac{1-f(\epsilon_{R\mathbf{k}_1})}{\epsilon_{R\mathbf{k}_1}-
    \omega+E_n-E_m} \,
  \frac{1-f(\omega+E_m-E_l)}{1-f(\omega)}
- \frac{f(\epsilon_{R\mathbf{k}_1})}
  {\epsilon_{R\mathbf{k}_1}-\omega+E_l-E_n}
  \frac{1-f(\omega+E_m-E_l)}{1-f(\omega)}
  \bigg\}.
\label{appB.Sigma3.7}
\end{align}
\end{widetext}
Finally, the sum over $\mathbf{k}_1$ leads to Eq.~(\ref{selfenergy2nd3rd})
for the self-energy. Here we assume $x\gg|\epsilon_d|\gg\omega,E_n$ for all states
$n$ and only keep the terms that diverge at $\omega=E_n-E_m$
and low temperatures.

\end{document}